\documentclass{jfm}
\usepackage{graphicx}
\usepackage{epstopdf,epsfig}
\usepackage{amsmath}
\usepackage{mathtools}
\usepackage{tikz}
\usepackage{comment} 
\usepackage{subcaption}
\usepackage{siunitx}
\usepackage{units}
\usepackage{soul}
\usepackage{ulem} 
\usepackage{multirow}

\newcommand{\reg}{\mathrm{reg}}
\newcommand{\s}{\mathsf{s}}
\newcommand{\LL}{\textsf{L}}
\newcommand{\HH}{\textsf{H}}
\newcommand{\II}{\textsf{I}}
\newcommand{\TT}{\textsf{T}}

\renewcommand{\d}{\mbox{d}}
\providecommand\bnabla{\boldsymbol{\nabla}}
\newcommand{\uu}{\textbf{u}}
\newcommand{\xx}{\textbf{x}}

\shorttitle{Sustained stratified shear flows}
\shortauthor{A. Lefauve and others}

\title{Regime transitions and energetics of sustained stratified shear flows}

\author{Adrien Lefauve\aff{1}, J. L. Partridge\aff{1}, P. F. Linden\aff{1}}

\affiliation{\aff{1}Department of Applied Mathematics and Theoretical Physics, Centre for Mathematical Sciences, Wilberforce Road, Cambridge CB3 0WA, UK}

\begin{document}

\maketitle

\begin{abstract}

We describe the long-term dynamics of sustained stratified shear flows in the laboratory. The Stratified Inclined Duct (SID) experiment sets up a two-layer exchange flow in an inclined  duct connecting two reservoirs containing salt solutions of different densities. This flow is primarily characterised by two non-dimensional parameters: the tilt angle of the duct with respect to the horizontal, $\theta$  (a few degrees at most), and the Reynolds number $Re$,  an input parameter based on the density difference driving the flow. The flow can be sustained with constant forcing over arbitrarily long times and exhibits a wealth of dynamical behaviours representative of geophysically-relevant sustained stratified shear flows. Varying $\theta$ and $Re$ leads to four qualitatively different regimes: laminar flow; mostly laminar flow with finite-amplitude, travelling Holmboe waves; spatio-temporally intermittent turbulence with substantial interfacial mixing; and sustained, vigorous interfacial turbulence (Meyer \& Linden, \emph{J. Fluid Mech.}, vol. 753, 2014, pp. 242--253). We seek to explain the scaling of the transitions between flow regimes in the two-dimensional plane of input parameters $(\theta, Re)$. We improve upon previous studies of this problem by providing a firm physical basis and non-dimensional scaling laws that are mutually consistent and in good agreement with the empirical transition curves we inferred from 360 experiments spanning $\theta \in [-1^\circ,6^\circ]$ and $Re\in [300,5000]$. To do so, we employ  state-of-the-art simultaneous volumetric measurements of the density field and the three-component velocity field, and analyse these experimental data using time- and volume-averaged potential and kinetic energy budgets. We show that regime transitions are caused by an increase in the non-dimensional time- and volume-averaged kinetic energy dissipation within the duct, which scales with $\theta Re$ at high enough angles. As the power input scaling with $\theta Re$ is increased above zero, the two-dimensional, parallel-flow dissipation (power output) increases to close the budget through an increase in the magnitude of the exchange flow, incidentally triggering Holmboe waves above a certain threshold in interfacial shear. However, once the hydraulic limit of two-layer exchange flows is reached, two-dimensional dissipation plateaus and three-dimensional dissipation at small scales (turbulence) takes over, first intermittently, and then steadily, in order to close the budget and follow the $\theta Re$ scaling.
This general understanding of regime transitions and energetics in the SID experiment may serve as a basis for the study of more complex sustained stratified shear flows found in the natural environment.

\end{abstract}



\section{Introduction}\label{sec:intro}

Turbulence is still an `unsolved problem', and the stabilising buoyancy forces that characterise \emph{stratified turbulence} add further complexity. The spatio-temporal scales involved in the physics of (stratified) turbulent flows make them very difficult to understand with our current computational capabilities and brain power. 

The historical and dominant angle of attack to this problem is to attempt to model the `small-scale' (inaccessible) physics of turbulence and mixing using the `large-scale' (accessible) properties of the flows. A much-pursued goal is the ability, for any given flow, to predict its regime (e.g. laminar, intermittently turbulent, fully turbulent), rate of energy dissipation and mixing efficiency (so-called `outputs' variables) using only a small number of `input' non-dimensional parameters characterising the flow (for four decades of reviews on mixing efficiency, see e.g. \cite{linden_mixing_1979,fernando_turbulent_1991,ivey_density_2008,gregg_mixing_2018}). Drawing on the power of dimensional analysis and theoretical scaling laws, the hope is that empirical relationships obtained under controlled conditions can then be extrapolated beyond laboratory or simulated scales. Following this tradition of research, the aim of this paper is the quantitative study of flow regimes, and particularly of the transitions between them, from a non-dimensional perspective.  

Stably stratified shear flows are a class of flows particularly relevant to the environment. Many of these flows are \emph{sustained} over long periods of time through quasi-steady forcing: for example exchange flows in straits, estuaries (e.g. \cite{geyer_mixing_2010}), coastal inlets (e.g. \cite{farmer_stratified_1999}), deep ocean overflows (e.g. \cite{van_haren_extremely_2014}) and stratified flows in the atmospheric boundary layer (e.g. \cite{mahrt_stably_2014}). In this paper, we address these general and geophysically-relevant sustained stratified shear flows using a simple laboratory experiment: the Stratified Inclined Duct experiment.


\section{The Stratified Inclined Duct (SID) experiment}

We introduce the experiment central to this paper in \S~\ref{sec:setup}, and our experimental measurements and flow regime visualisations in \S~\ref{sec:measurements}. We discuss the distribution of flow regimes in the space of input parameters and motivate this paper by reviewing the most relevant literature in \S~\ref{sec:intro-reg-bif}. We then build on the previous sections to reformulate the above aim in more specific terms and outline the paper in \S~\ref{sec:outline}.

\subsection{Setup, notation and non-dimensionalisation} \label{sec:setup}

\begin{figure}
\centering
\includegraphics[width=0.8\textwidth]{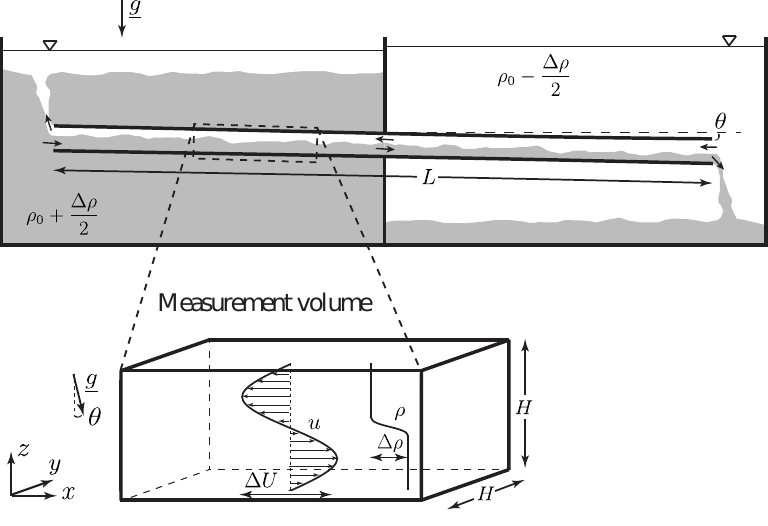}
    \caption{Schematics of the Stratified Inclined Duct (SID) experiment. The measurement volume inset shows the coordinate system and the notation used in this paper (in dimensional units). Note that the $x$ axis is aligned along the duct, resulting in gravity pointing at an angle $\theta$ from the $-z$ direction. Here, by definition, the duct is inclined at a positive angle $\theta>0^\circ$, resulting in a positive forcing of the flow by the streamwise projection of gravity $g\sin\theta>0$.}
    \label{fig:setup}
\end{figure}

The Stratified Inclined Duct experiment (hereafter abbreviated SID) is sketched in figure \ref{fig:setup}. This conceptually simple experiment consists of two reservoirs initially filled with aqueous salt solutions of different densities $\rho_0 \pm \Delta \rho/2$, connected by a long rectangular duct that can be tilted at a small angle $\theta$ from the horizontal (this is made possible by a flexible seal between the duct and the barrier separating the two reservoirs). At the start of the experiment, the duct is opened. After a brief transient gravity current, a two-layer exchange flow is sustained for long periods of time in the duct. This sustained stratified shear flow is the focus of this paper.

This flow is driven by two distinct forcing mechanisms: \emph{(i)} a horizontal hydrostatic pressure gradient of opposite sign in each layer, resulting from each end of the duct sitting in reservoirs containing fluids of different densities, which is present even when the duct is horizontal (i.e when $\theta=0^\circ$); \emph{(ii)} the gravitational acceleration of the buoyant layer upward (to the left) and the dense layer downward (to the right) when the tilt angle is positive $\theta>0^\circ$, defined here by the duct being raised in the denser reservoir, as shown in figure~\ref{fig:setup}. The relative influence of these two forcing mechanisms  will be discussed in \S~\ref{sec:two-layer-model}.

To the authors' knowledge, the SID experiment was first studied by \cite{macagno_interfacial_1961}. It was independently `rediscovered' by \cite{kiel_buoyancy_1991} and more recently by \cite{meyer_stratified_2014} (hereafter ML14), who coined the name. ML14 correctly recognised that the two-layer exchange flow was maximal because it is hydraulically controlled at both ends of the duct where it meets the reservoirs through a sharp change in geometry (an idea already present in \cite{wilkinson_buoyancy_1986}). In other words, the flow is subcritical with respect to long interfacial waves inside the duct, and critical at either end, preventing the propagation of information (in particular of the exchange flow rate) from the exterior into the duct (see ML14, \citealp[\S~3]{lefauve_structure_2018}, and \citealp[\S~1.3.2]{lefauve_waves_2018} for more details). The exchange flow is sustained in a quasi-steady state until the controls are `flooded' by the accumulation of fluid of a different density coming from the other reservoir. With each reservoirs holding approximately 100~l of fluid in our current setup, a typical experiment can last several minutes, which represents many duct transit times.

Our notation is shown in the measurement volume inset in figure \ref{fig:setup} and follows that of \cite{lefauve_structure_2018} (hereafter LPZCDL18). The duct considered in this paper has length $L=1350$~mm and a square cross-section of $H=45$~mm (the same dimensions as LPZCDL18 but smaller than ML14). The streamwise $x$ axis is aligned along the duct and the spanwise $y$ axis across the duct, making the $z$ axis tilted at an angle $\theta$ from the vertical (resulting in a non-zero streamwise projection of gravity $g\sin\theta$ providing the gravitational forcing). All coordinates are centred in the middle of the duct, such that $-L/2 \leq x \leq L/2$ and $-H/2 \leq y,z \leq H/2$. The velocity vector field has components $\uu(x,y,z,t)=(u,v,w)$ along $x,y,z$, and we denote the density field by $\rho(x, y, z, t)$. 

The parameters believed to play important roles are the geometrical parameters: $L$, $H$, $\theta$, and the dynamical parameters: the reduced gravity $g' \equiv g \Delta \rho/\rho_0$ (under the Boussinesq approximation of small density differences $0<\Delta \rho / \rho_0 \ll 1 $),  the kinematic viscosity of water $\nu = 1.05\times 10^{-6}$~m${}^2$~s$^{-1}$ and the molecular diffusivity of salt $\kappa_s = 1.50\times 10^{-9}$~m${}^2$~s$^{-1}$. From these six parameters having two dimensions (of length and time), we construct four independent non-dimensional parameters below.

In this maximal exchange flow, the velocity scale $\Delta U$ is not an independent parameter; it is primarily set by the phase speed of long interfacial gravity waves. To understand this, we follow the literature (see e.g. \cite{armi_hydraulics_1986,lawrence_hydraulics_1990}) and define the composite Froude number of this two-layer flow as
\begin{equation}\label{composite-Froude}
G^2(x) \equiv F_1^2(x) + F_2^2(x), \quad \textrm{where} \quad F_i^2(x) \equiv \frac{\langle u_i^2(x) \rangle_{y,z_i}}{g'h_i(x)}
\end{equation}
is the Froude number of layer $i$, $\langle \cdot \rangle_{y,z_i}$ denotes spanwise and vertical averaging over the depth $h_i$ of each layer, and the symbol $\equiv$ denotes a definition.  
In the idealised case of frictionless, horizontal ducts ($\theta =0^\circ$), the flow is streamwise invariant and $G$ takes everywhere the value at the centre of the duct
\begin{equation}\label{G0}
G(x) = G(0) =  2\frac{\langle |u| \rangle_{y,z}}{\sqrt{g'H}}, 
\end{equation}
where $\langle \cdot \rangle_{y,z}$ denotes averaging over the whole duct cross section. The second equality results from \eqref{composite-Froude} and the symmetry of the flow at $x=0$ guaranteed by the Boussinesq approximation ($\langle |u_1| \rangle_{y,z}=\langle |u_2| \rangle_{y,z}$ and $h_1=h_2=H/2$). Note that here and in the remainder of the paper, we assume that the exchange flow has zero \emph{net} (or `barotropic') flow rate, i.e.
\begin{equation}\label{no-barotropic}
\langle u \rangle_{y,z}=0, 
\end{equation}
which is a good approximation in the present setup.
Hydraulic control requires that $G^2 \equiv 1$ \citep{armi_hydraulics_1986}, which gives the following layer-averaged velocity
\begin{equation}\label{layer-avgd-vel}
\langle |u| \rangle_{y,z} = \frac{\sqrt{g'H}}{2}. 
\end{equation}
With the addition of viscous friction and/or of a non-zero tilt angle, the flow is no longer streamwise invariant: $G(x)$ is maximal at the ends ($x=\pm L/2$) and minimal in the centre ($x=0$). Since the criticality condition is imposed at the ends where the controls occur $G(\pm L/2)=1>G(0)$, the velocity scale $\langle |u| \rangle_{y,z} = (\sqrt{g'H}/2)G(0)$ is lower than the inviscid upper bound \eqref{layer-avgd-vel} that we call `hydraulic limit' (see \cite{gu_analytical_2005} for more details). As first observed in ML14 (see their figure~7) and as we shall substantiate in \S~\ref{sec:two-layer-model}, this hydraulic limit is however generally achieved when a positive tilt angle $\theta > 0^\circ$ is added to counterbalance the dissipative effects of viscosity.

Due to the moderate Reynolds numbers and the long duct investigated in the present setup, the velocity profiles are usually significantly affected by viscosity in the sense that viscous boundary layers at the walls and interface are partially or fully developed. Generally, we find that the peak velocities in each layer are at most around twice the layer-averaged values corresponding to the hydraulic limit \eqref{layer-avgd-vel}, i.e. $\max_{y,z} |u| \approx 2 \langle |u|\rangle_{y,z} \approx  \sqrt{g'H}$. We choose to non-dimensionalise velocities by this characteristic `peak' value, i.e. half the total (peak-to-peak) velocity jump $\Delta U$ (shown in the inset in figure~\ref{fig:setup})
\begin{equation}\label{definition-DeltaU}
\frac{\Delta U}{2} \equiv \sqrt{g'H}.
\end{equation}
We thus define the non-dimensional velocity vector as $\tilde{\uu} \equiv \uu /(\Delta U/2)$ such that in general $-1 \lesssim \tilde{u} \lesssim 1$ (noting that the streamwise velocity is dominant in this flow, i.e. $|\tilde{u}| \gg |\tilde{v}| ,|\tilde{w}|$). For consistency, we choose $H/2$ as the length scale, defining the non-dimensional position vector as $\tilde{\xx} \equiv \xx/(H/2)$ such that $-1 \le \tilde{y},\tilde{z} \le 1$, and $-A \le \tilde{x} \le A$, where  the aspect ratio of the duct is
\begin{equation}\label{definition-A}
A \equiv \frac{L}{H}.
\end{equation}
Consequently, we non-dimensionalise time by the advective time unit $H/\Delta U = 1/(2\sqrt{g'/H})$: $\tilde{t}\equiv 2\sqrt{g'/H}t$ (hereafter abbreviated ATU). The dimensionless density field is defined as $\tilde{\rho} \equiv (\rho - \rho_0)/(\Delta \rho /2)$, such that  $-1 \le \tilde{\rho} \le 1$. 

Using the previously defined velocity and length scales, we construct the Reynolds number
\begin{equation} \label{definition_Re}
Re \equiv \frac{\frac{\Delta U}{2} \frac{H}{2}}{\nu} = \frac{\sqrt{g'H}H}{2\nu} = 1.
42\times 10^4 \, \sqrt{\frac{\Delta \rho}{\rho_0}},
\end{equation}
where the last equality shows that $Re$ is a function of the driving density difference $\Delta \rho/\rho_0$ alone (the prefactor only holds for aqueous salt solutions in the geometry investigated here). In this paper, we present experiments in the range $\Delta \rho/\rho_0 \in [5\times 10^{-4}, 1.3\times10^{-1}]$, i.e. $Re \in [300, 5000]$.

The criticality condition $G^2=1$ adds another dimensional parameter, $\Delta U$, to our previous set of six input parameters. This velocity scale set by the criticality of the exchange flow can be recast as an overall Richardson number, expressed as the non-dimensional product of the density, length and inverse square velocity scales, and which here takes a constant value 
\begin{equation} \label{definition_Ri}
Ri \equiv \frac{\frac{g}{\rho_0}\frac{\Delta \rho}{2} \frac{H}{2}}{\Big(\frac{\Delta U}{2}\Big)^2} = \frac{1}{4},
\end{equation}
by definition of $\Delta U$ in \eqref{definition-DeltaU}. 

Our last non-dimensional parameter is the Schmidt number, the ratio of the momentum to salt diffusivity
\begin{equation} \label{definition_Sc}
Sc\equiv \frac{\nu}{\kappa_s}.
\end{equation}

In summary, we have a total of four free independent non-dimensional input parameters: $\theta$, $A$, $Re$, $Sc$, and one imposed parameter $Ri$. For the apparatus considered, we have $A=30$, $Sc=700$, $Ri=1/4$, and we have the freedom to vary $\theta$ and $Re$ (by varying $\Delta \rho/\rho_0$), allowing us access to a wide range of flow regimes as ML14 demonstrated and as we show in \S~\ref{sec:measurements}-\ref{sec:intro-reg-bif}. 
Henceforth, we drop the tildes and, unless explicitly stated otherwise, use non-dimensional variables throughout.

\subsection{Measurements} \label{sec:measurements}

In this section we introduce the three types of experimental measurements discussed in this paper: shadowgraph; mass flux; and volumetric three-dimensional, three-component (3D-3C) measurements of the velocity and density fields. We then discuss 3D-3C visualisations of flows in each regime to highlight some key features.

\subsubsection{Shadowgraph} \label{sec:SG}

Shadowgraph observations of the flow in the duct were employed by ML14 to identify and classify four qualitatively different flow regimes depending on $\theta$ and $Re$ (see their figure~3) that they called $\LL$, $\HH$, $\II$ and $\TT$:
\begin{itemize}
\item \LL \ : \textbf{L}aminar steady flow, with a thin, flat density interface between the two counter-flowing layers; 
\item \HH \ : mostly laminar flow, with finite-amplitude \textbf{H}olmboe waves propagating on the interface;
\item \II \ : spatio-temporally \textbf{I}ntermittent turbulence with small-scale structures and mixing that are conspicuous in the shadowgraph; 
\item \TT \ :  steadily sustained \textbf{T}urbulence with significant small-scale structures and a thick interfacial mixing layer. 
\end{itemize}
In this paper, we followed ML14 and carried out similar shadowgraph observations in our setup to classify hundreds of observed flows into these four qualitative regimes: (note that they are identical to those described in \cite{macagno_interfacial_1961}, despite ML14 not being aware of their work).

\subsubsection{Mass flux} \label{sec:MF}

We first define the instantaneous `volume flux' $Q>0$, or exchange volume flow rate defined as the duct-averaged absolute value of the streamwise velocity:
\begin{equation}\label{Q-def}
Q(t) \equiv \langle |u| \rangle_{x,y,z},
\end{equation}
where we recall that we assume no net flow, i.e. $\langle u \rangle_{x,y,z}=0$. 

By analogy, we define the instantaneous `mass flux' $Q_m>0$, or exchange mass flow rate as
\begin{equation}\label{Qm-def}
Q_m(t) \equiv \langle \rho u \rangle_{x,y,z}
\end{equation}
The $x$-averaging in these definitions is, strictly-speaking, unnecessary by conservation of volume along the duct, but we retain it as it will be employed to reduce experimental noise when evaluating $Q$ and $Q_m$ using three-dimensional data of $u(x,y,z)$ later. Note that $Q_m=Q$ in the absence of net flow and mixing (since in this case $\rho=\mathrm{sgn}(u)$), but in general $0<Q_m<Q$ in the presence of mixing (the distribution of $\rho$ is no longer bimodal and becomes continuous). 

In a subset of the experiments in which shadowgraph observations were made, we also carried out mass flux measurements of $\langle Q_m\rangle_t$, where $\langle \cdot \rangle_t$ denotes averaging over the length of an experimental run. They were carried out as in ML14 using salt mass balances, i.e. by measuring the mean density of the solutions in each reservoir ($\rho_0 \pm \Delta \rho/2$) before and after the experiment (for more details see \citealp[\S~2.2]{lefauve_waves_2018}, hereafter L18). Note the relation between our time-averaged mass flux and its equivalent definition in ML14, who called it the `Froude number' $F\equiv  \sqrt{2} \langle Q_m \rangle_t$. 

The hydraulic limit for the volume flux set by the maximal exchange flow condition \eqref{layer-avgd-vel} can be rewritten in non-dimensional form as $Q=0.5$ (non-dimensionalising $u$ by \eqref{definition-DeltaU}). We therefore have in general
\begin{equation}\label{Q-inequalities}
0< Q_m \le Q \le 0.5.
\end{equation}
The first two inequalities always hold by definition whereas the last inequality is the theoretical hydraulic limit that does not always hold in the experiments (we occasionally measured up to $Q_m \approx 0.6$).

\subsubsection{Volumetric three-dimensional, three-component (3D-3C) measurements} \label{sec:3d-3c}

To provide a quantitative basis to the qualitative shadowgraph observations and subsequent categorisation into flow regimes, we investigate in this paper the detailed energetics underpinning each regime.  To do so, we employed simultaneous measurements of the density field and three-dimensional, three-component (3D-3C) velocity field in a volume, as sketched in the inset of figure~\ref{fig:setup}. 

These measuremements relied on a novel technique introduced by \cite{partridge_new_2018} in which a thin, pulsed vertical laser sheet (in the $x-z$ plane) is scanned rapidly back and forth in the spanwise direction (along $y$) to span a duct subvolume of non-dimensional cross-section $2\times 2$ and non-dimensional length $\ell$ (typically a small fraction of full duct length $\ell \ll 2A$). Simultaneous stereo Particle Image Velocimetry (sPIV) and Planar Laser Induced Fluorescence (PLIF) are employed to obtain the three-dimensional, three-component velocity and density fields $(u,v,w,\rho)(x,y_i,z,t_i)$ in successive $x-z$ planes at spanwise locations $y=y_i$ and respective times $t=t_i$. Three-dimensional volumes containing $n_y$ planes (i.e. $i=1,2,\cdots,n_y$) are then reconstructed from these plane measurements.
These volumetric 3D-3C measurements are only \emph{near-instantaneous} in the sense that each plane $(x,y_i,z,t_i)$ is separated from the previous one by a small time increment $\delta t \equiv t_{i}-t_{i-1}$, resulting in each volume being constructed over a non-dimensional time $\Delta t \equiv n_y \delta t$.  The experimental protocol and details to obtain the measurements used in this paper are identical to those discussed in LPZCDL18~\S~3.3-3.4. 

This technique provides high-resolution measurements of $(u,v,w,\rho) (x,y,z,t)$ with a typical number of data points in each coordinate $(n_x,n_y,n_z,n_t) \approx (500,30,100,300)$ per experiment (after processing 150~GB of raw data). The details of the volume location $\bar{x}$, length $\ell$, duration of an experiment $\tau$, and resolution $(\Delta x, \Delta y, \Delta z, \Delta t) \equiv (\ell/n_x, 2/n_y,2/n_z,\tau/n_t)$ for all 3D-3C experiments discussed in this paper will be given in \S~\ref{sec:exp-validation} (table~\ref{tab:3d-3c-expts}). We discuss the physical constraints setting bounds on all of the above values in appendix \ref{sec:appendix-exp}.

Finally, we enforced incompressibility in all the measured volumetric 3D-3C velocity fields by imposing $\bnabla \cdot \uu =0$ for each of the $n_t$ volumes. We employed the recent weighted divergence correction scheme of \cite{wang_weighted_2017}, which constitutes an improved and much faster variant of the general algorithm of \cite{de_silva_minimization_2013}. Encouragingly, we found that the level of correction needed (the volume-averaged relative $L^2$ distance between the original and corrected fields) was typically small (at most a few \%).

\subsubsection{Flow regime visualisations} \label{sec:reg-vis}

\begin{figure}
\centering
\includegraphics[width=\textwidth]{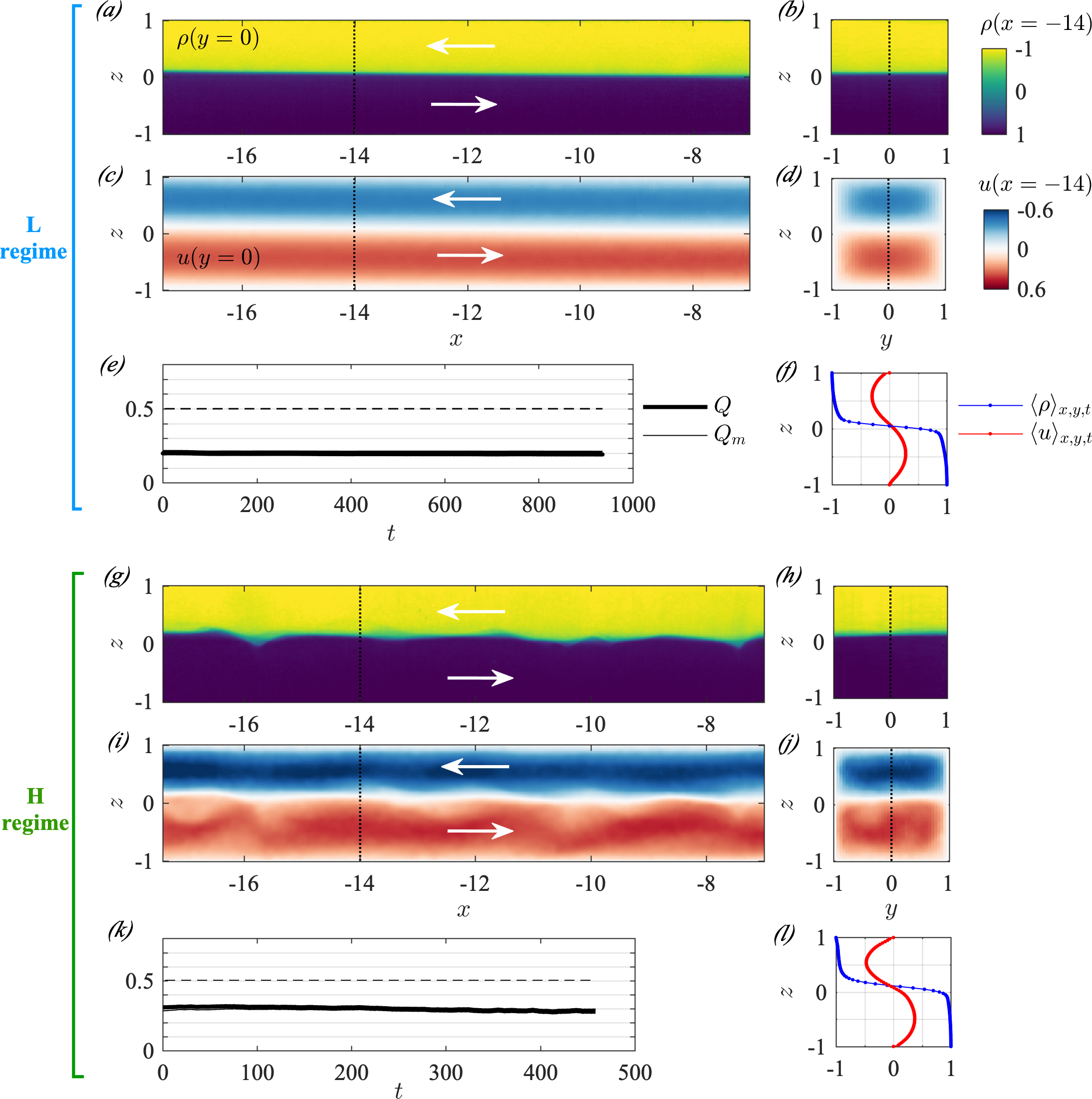}
    \caption{Comparative visualisations of a typical  \emph{(a-f)} $\LL$ flow ($\theta = 2^\circ$, $Re = 398$) and \emph{(g-l)} $\HH$ flow ($\theta = 1^\circ$, $Re = 1455$). The $\II$ and $\TT$ regimes are shown in figure~\ref{fig:regime-snaps-I-T}. The $\LL$ and $\HH$ data correspond respectively to experiments L1 and H1 listed in table~\ref{tab:3d-3c-expts} (discussed later). For each experiment, we plot the density field $\rho$ and streamwise velocity field $u$ in \emph{(a,c,g,i)}  the vertical mid-plane of the volume $y=0$, and in \emph{(b,d,h,j)} the arbitrary cross-sectional plane $x=-14$, all for a single arbitrary temporal snapshot: $t=150$ in \emph{(a-d)}, and $t=261$ in \emph{(g-j)}. Colour bars are identical for all plots showing density or velocity and are thus not repeated. Dotted vertical lines in the $y=0$ plane \emph{(a,c,g,i)} indicate the location of the $x=-14$ plane in \emph{(b,d,h,j)} and conversely. White arrows indicate the direction of the flow in each layer (in agreement with the notation of \S~\ref{sec:setup} and figure~\ref{fig:setup}). In addition, we plot for each experiment:  \emph{(e,k)} the temporal evolution of the volume flux $Q(t)$ and mass flux $Q_m(t)$ (the dashed line is the hydraulic limit $Q=0.5$); and  \emph{(f,l)} the mean vertical density and streamwise velocity profiles (the dot symbols indicate the vertical resolution of the data).  } \label{fig:regime-snaps-L-H}
\end{figure}

\begin{figure}
\centering
\includegraphics[width=\textwidth]{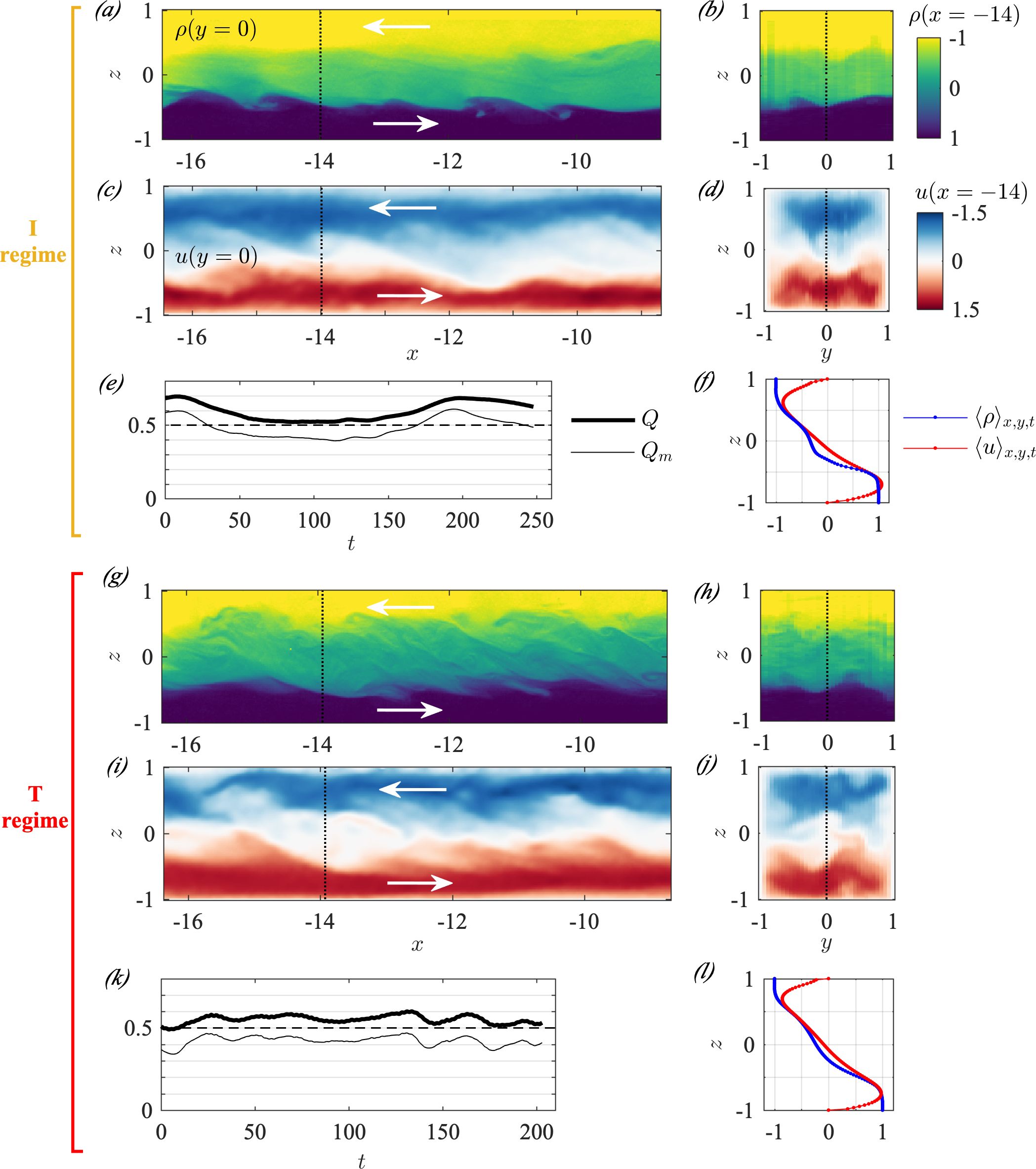}
    \caption{Comparative visualisations of a typical \emph{(a-f)} $\II$ flow ($\theta = 6^\circ$, $Re = 777$) and  \emph{(g-l)} $\TT$ flow ($\theta = 6^\circ$, $Re = 1256$), corresponding respectively to experiments I4 and T2 of table~\ref{tab:3d-3c-expts} (discussed later). The legend is identical to that of figure~\ref{fig:regime-snaps-L-H}, except for the temporal snapshots used here: $t=55$ in \emph{(a-d)} and $t=168$ in \emph{(g-j)}. } \label{fig:regime-snaps-I-T}
\end{figure}

We show visualisations of a flow characteristic of each of the four regimes in figure~\ref{fig:regime-snaps-L-H} ($\LL$ and $\HH$ regimes) and figure~\ref{fig:regime-snaps-I-T} ($\II$ and $\TT$ regimes). We used the 3D-3C measurements described above to plot, for each regime, the same three types of data for side-by-side comparison:
\begin{itemize}
    \item instantaneous snapshots of the density field $\rho$ and streamwise velocity field $u$ in the vertical mid-plane $y=0$ of the measurement volume (`top left' two panels \emph{a,c,g,i}), and in the arbitrary cross-sectional plane $x=-14$ (`top right' two panels \emph{b,d,h,j});
    \item time series of the volume flux $Q(t)$ and mass flux $Q_m(t)$ (`bottom left' panels \emph{e,k});
    \item averaged vertical density profile  $\langle \rho \rangle_{x,y,t}(z)$ and velocity profile $\langle u \rangle_{x,y,t}(z)$ (`bottom right' panels \emph{f,l}).
\end{itemize}
For more complete visualisations, including horizontal planes and the other velocity components $v$ and $w$ (not shown here), see \cite{partridge_new_2018}.


We observe that the $\LL$ and $\HH$ flows have a sharp density interface with a tanh-like vertical profile (figure~\ref{fig:regime-snaps-L-H}\emph{(a,b,f,g,h,l)}), while the $\II$ and $\TT$ flows have a mixing layer  (figure~\ref{fig:regime-snaps-I-T}\emph{(a,b,f,g,h,l)}), i.e. a central layer in which the vertical density gradient is smaller than the values immediately above and below it as a result of turbulent mixing across the interface. 

In the $\LL$ and $\HH$ regimes, the streamwise velocity profile has a sine-like vertical structure (figure~\ref{fig:regime-snaps-L-H}\emph{(f,l)}) indicative of fully-developed velocity boundary layers (expected when $Re\lesssim 50A=1500$). By contrast, in the $\II$ and $\TT$ regimes, interfacial turbulence creates a region of approximately constant velocity gradient across the mixing layer and `pointier' maxima that are pushed closer to the top and bottom walls (figure~\ref{fig:regime-snaps-I-T}\emph{(f,l)}) especially when turbulence is more intense and sustained in the $\TT$ flow. 

We also note that the $\LL$ flow is largely \emph{(i)} parallel, i.e. independent of the streamwise direction $x$, except for a very slight downward slope of the interface typical of such flows (discussed later in \S~\ref{sec:two-layer-model}); \emph{(ii)} steady in time; \emph{(iii)} symmetric about the $y=0$ and $z=0$ planes. By contrast, the $\HH$ flow breaks the $x$- and $t$-invariance with a set of travelling, symmetric Holmboe waves distorting the density and velocity interfaces in a characteristic `cusp'-like pattern and in a quasi-periodic fashion (these `confined Holmboe waves' were the focus of LPZCDL18). In addition,  complex three-dimensional wave motions in the velocity field  break the $y=0$ and $z=0$ symmetries (figure~\ref{fig:regime-snaps-L-H}\emph{(i,j)}). 

In the $\II$ and $\TT$ flows, the departure from both the $x,t$ invariances and the $y,z=0$ symmetries at any instant in time is even greater, owing to large, three-dimensional turbulent fluctuations (figure~\ref{fig:regime-snaps-I-T}). Based on the deflections in the position of the density and velocity interfaces, the spatial scales of these fluctuations, and the amplitude of the temporal fluctuations in the $Q(t)$ and $Q_m(t)$ time-series, it is tempting to classify the $\LL$ and $\HH$ flows in one group based on their similarity, and the $\II$ and $\TT$ regimes in a different group. The $\LL-\HH$ flows have lower volume and mass flux, which are equal in the absence of mixing ($Q_m\approx Q \approx 0.2-0.3$), while the $\II-\TT$ flows have higher fluxes and significant mixing ($Q_m \approx 0.4-0.5 < Q \approx 0.5-0.6$, close to the hydraulic limit). 

Large temporal fluctuations in both $Q$ and $Q_m$ are  observed in the $\II$ and $\TT$ regimes, but $\II$ flows tend to exhibit a component with longer pseudo-period associated with oscillations between laminar and turbulent events (sometimes in a quasi-periodic fashion with period $O(100$~ATU$)$). This is visible in the $\II$ flow here (figure~\ref{fig:regime-snaps-I-T}\emph{(e)}): the start of a turbulent event  (shown here in the snapshots figure~\ref{fig:regime-snaps-I-T}\emph{(a-d)} at $t=55$) follows the instability of an accelerating, largely laminar, three-layer flow. A peak in the volume flux at $t \approx 10$ triggered large-amplitude waves at both density interfaces which started overturning at $t\approx 40$ and initiated a turbulent event slowing down the flow (decreasing $Q$ and $Q_m$). Relaminarisation followed at $t\approx 130$ (increasing $Q$ and $Q_m$), and another cycle started (note that only one cycle was recorded here). 

The basic characteristics of flow regimes described above are summarised in table~\ref{tab:regimes}.
\begin{table}
  \begin{center}
\def~{\hphantom{0}}
\setlength{\tabcolsep}{6pt}
  \begin{tabular}{lcccc}
     & $\quad \LL \quad$ & $\quad \HH \quad$ & $\quad \II \quad$ & $\quad \TT \quad$ \\ [5pt]
\multirow{1}{*}{\begin{tabular}{l} Invariance in $x,t$\end{tabular}} & $\checkmark$ & $\sim$ & $\times$  & $\times$ \\ [5pt] 
\multirow{1}{*}{\begin{tabular}{l} Symmetry about $y,z=0$ \end{tabular}} & $\checkmark$ & $\sim$ & $\times$  & $\times$  \\ [5pt] 
\multirow{1}{*}{\begin{tabular}{l} Large $Q, Q_m \approx 0.5 $\end{tabular}}  & $\times$ & $\times$  & $\checkmark$ & $\checkmark$ \\ [5pt] 
\multirow{1}{*}{\begin{tabular}{l} Interfacial mixing \end{tabular}}    &  $\times$ & $\times$  & $\checkmark$ & $\checkmark$ \\ [5pt] 
\multirow{1}{*}{\begin{tabular}{l} Small spatial scales \end{tabular}} &  $\times$ & $\sim$  & $\checkmark$ & $\checkmark$ \\ [5pt] 
\multirow{2}{*}{\begin{tabular}{l} Laminar-turbulent \\ [0pt] periodicity \end{tabular}}  &  $\times$ & $\times$  & $\checkmark$ & $\times$ \\ [5pt] 
\end{tabular} 
    \caption{Basic characteristics of flow regimes inferred from figures \ref{fig:regime-snaps-L-H}-\ref{fig:regime-snaps-I-T}.  Symbol $\sim$ indicates a relatively small effect.}
\label{tab:regimes}
  \end{center}
\end{table}

\subsection{Regime diagram and previous studies} \label{sec:intro-reg-bif}

\subsubsection{Regime diagram}

\begin{figure}
\centering
\includegraphics[width=0.8\textwidth]{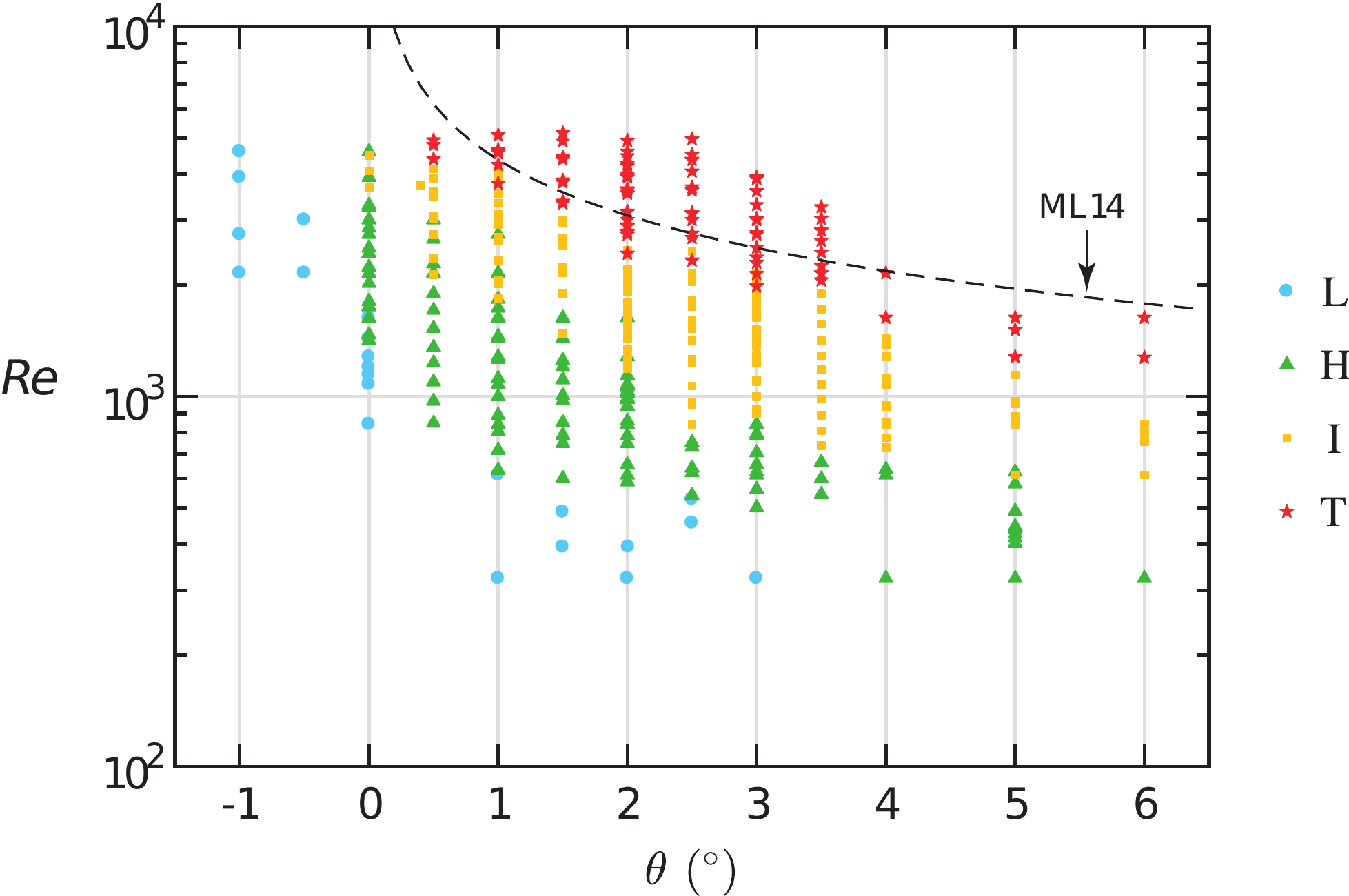}
    \caption{Regime diagram in the $(\theta, \, Re)$ plane of non-dimensional input parameters totalling 360 data points. In dashed, the $\II\rightarrow \TT$ transition curve obtained from experiments in a larger duct by ML14 (see \S~\ref{sec:ML14}). } \label{fig:regime-data-input-params}
\end{figure}

The map of flow regimes \LL, \HH, \II, \TT \, in the $\theta-Re$ plane of input parameters is shown in figure~\ref{fig:regime-data-input-params}.
This \emph{regime diagram} features a total of 360 points, corresponding to the qualitative identification of regimes for 360 $(\theta, Re)$ couples. Out of these, 312 were determined from shadowgraph observations (\S~\ref{sec:SG}) as in ML14, 35 were determined from 3D-3C experiments (\S~\ref{sec:3d-3c}), and 13 from simpler planar PIV and PLIF measurements (two-dimensional, two-component, in the $y=0$ plane) that were carried out before the 3D-3C system was operational (these measurements are not discussed in this paper). 

We observe that the \LL, \HH, \II \, and \TT \, regimes largely occupy distinct regions of the $\theta-Re$ plane, with little overlap. We refer to the boundaries between each regime respectively as the $\LL\rightarrow\HH$, $\HH\rightarrow\II$, and $\II\rightarrow\TT$ transitions, which can be described by simple open curves in the $\theta-Re$ plane. To fix ideas, we may formally define a `regime function' $\reg$ taking arbitrary but increasing values such as
\begin{equation} \label{def-reg}
  \reg \equiv  
  1 \ \textrm{for} \ \LL, \ \  2 \ \textrm{for} \ \HH, \ \  3 \ \textrm{for} \ \II, \ \ 4 \ \textrm{for} \ \TT.
\end{equation}
Finding the scaling of flow transitions is equivalent to finding the functional dependence of the regime function with respect to the two input parameters varied in this paper: $\reg(\theta,Re)$. Such `transition curves' can then be described, for example, by the equations $\reg=1.5,\, 2.5, \, 3.5$.  

Sufficiently far from the transitions curves, the flow regime is a repeatable characteristic of the experiment (and of the underlying dynamical system) for a choice of input parameters $(\theta,\,Re)$. The slight overlap between regimes near the transitions is interesting, and may be explained by two potential reasons:
\begin{enumerate}
\item the flow regime may not be a reproducible characteristic of the experiment (and of the underlying dynamical system) near the transitions due to its sensitivity to flow parameters, and/or to initial conditions (the initial transients resulting from the way the experiment is started, which cannot be controlled accurately);

\item the qualitative (visual) identification of flow regimes, i.e. the very definition of `flow regime' is not appropriate near the transitions (i.e. not fine or consistent enough) to classify the flow into the four discrete categories of ML14.
\end{enumerate}

Note that throughout this paper, we use the term `regime \textit{transition}' to refer to the change in the qualitative long-term (asymptotic) dynamics of the flow caused by changes in the input parameters. Although mathematically such behaviour is typically referred to as a \textit{bifurcation}, we chose to avoid this term in this paper since we do not prove nor imply that the underlying dynamical system indeed exhibits strict bifurcations. This question is interesting but outside the scope of this paper.

\subsubsection{\cite{meyer_stratified_2014}} \label{sec:ML14}

The regime diagram in figure~\ref{fig:regime-data-input-params} complements that of ML14 (their figure~5). ML14 plotted it in the $\theta-\Delta \rho/(2\rho_0)$ plane) for 93 experiments using a larger duct ($H=100$~mm \emph{vs} $45$~mm) of the same aspect ratio ($A=30$). They sought an equation for the transition curves by arguing that, because of the presence of hydraulic controls (\S~\ref{sec:setup}), the kinetic energy in the flow was bounded by the scaling $(\Delta U)^2 \sim g'H$ (see \eqref{layer-avgd-vel} and \eqref{definition-DeltaU}) and thus it could not increase even in the presence of gravitational forcing when $\theta>0^\circ$. The dimensional `excess kinetic energy' $g'L\sin\theta$, gained by conversion from potential energy by the fluid travelling a distance $L$ along the duct in the streawise field of gravity $g'\sin \theta>0$, thus has to be dissipated by increased wave activity or turbulence. They non-dimensionalised this excess kinetic energy by $(\nu/H)^2$, thus forming the following Grashof number
\begin{equation} \label{grashof}
Gr \equiv \frac{g'L\sin\theta}{(\nu/H)^2} = 4 A  \sin\theta Re^2,
\end{equation}
where the first equality is their definition and the second equality uses our notation. They found reasonable agreement between this scaling in $\sin \theta Re^2$ (using two different aspect ratios $A=15,\,30$) and suggested the empirical equation $Gr=4\times10^7$ for the $\II\rightarrow\TT$ transition curve (see their figure~8). 

Their proposed $\II\rightarrow\TT$ transition curve is reproduced in dashed black in figure~\ref{fig:regime-data-input-params} (identified by the `ML14' arrow) to show that the agreement in our geometry (smaller duct) is less convincing. The ML14 curve lies entirely in the \TT \, region (i.e. it is `too high') and the discrepancy is particularly apparent at higher angles $\theta\gtrsim 4^\circ$ (which were not considered by ML14), suggesting that their proposed  `$\sin\theta Re^2$ scaling' of transitions may not be universal. 

\subsubsection{\cite{macagno_interfacial_1961}}

\cite{macagno_interfacial_1961} also mapped these same four regimes in a two-dimensional space (see their figure~8). However, instead of two input parameters such as $\theta$ and $Re$, they used a Froude number and a Reynolds number based on measured values of the actual (output) $\Delta U$ and of the vertical distance between the two maxima of $|u|$ (depth of the shear layer). They varied the tilt angle $\theta$ in non-trivial ways, sometimes during an experiment, in order to obtain target values of $\Delta U$ and therefore better control $Re$, and did not appear to realise the presence and importance of hydraulic controls (in fact, they may have disturbed them by their use of splitter plates at the ends of the duct). They recognised the importance of $Re$ in regime transitions, but not that of $\theta$, and were thus unable to propose a convincing physical model to substantiate the transitions.  

\subsubsection{\cite{kiel_buoyancy_1991}}
The third most relevant experimental study of regime transitions in the SID experiment is the (unpublished) PhD thesis of \cite{kiel_buoyancy_1991} (like most of the literature, he was not aware of \cite{macagno_interfacial_1961}).  Kiel proposed a heuristic scaling based on a `geometric Richardson number' $Ri_G \equiv (4A\tan \theta+16/9)^{-1}$ (using our notation). We interpret the parameter $Ri_G^{-1}$ as the non-dimensionalisation of the `excess kinetic energy' $g'L\sin\theta$ of ML14 by the actual kinetic energy of the hydraulically-controlled flow $(\Delta U)^2=g'H$, i.e. $Ri_G^{-1} \sim g'L\sin\theta/(g'H)=A\sin\theta$ (disregarding the additive constant $16/9$). Hence, when the excess energy to be dissipated becomes large compared with the maximum kinetic energy of the flow (high $Ri_G^{-1}$), transition to turbulence is expected. 

Contrary to \cite{macagno_interfacial_1961}, \cite{kiel_buoyancy_1991} only focused on the importance of $\theta$ on regime transitions, ignoring $Re$ which he (incorrectly) assumed large enough for viscous effects to be ignored. Although Kiel did use large $Re$ (of order $10^4$) using ducts of dimensions similar to that of ML14, the observations of ML14 at similar $Re$ highlighted the importance of the $Re$ scaling, which we substantiate in this paper. Consequently, his $Ri_G$ criterion, based a non-dimensionalisation of the excess kinetic energy by the velocity scale $\Delta U$ -- although apparently more physical than the somewhat arbitrary velocity scale $\nu/H$ of ML14 -- is fundamentally incapable of predicting regime transitions.

\subsection{Aim and outline} \label{sec:outline}

To summarise, we have seen that regime transitions in the SID depend on \emph{at least two input parameters}: $\theta$ and $Re$. The first two pioneering attempts to understand the transitions that we are aware of \citep{macagno_interfacial_1961,kiel_buoyancy_1991} each ignored one of them, proposing heuristic scalings based on (respectively) either $Re$ or $\theta$. More recently, ML14 correctly identified the $\theta-Re$ dependence, understood the consequence of hydraulic controls, and proposed a transition scaling following $Gr\sim \sin\theta Re^2=$~const. (see \eqref{grashof}). This scaling was based on heuristic arguments of `excess kinetic energy', which, as we will show this paper, are essentially correct but can be made more specific. However, the non-dimensionalisation by the square velocity scale $(\nu/H)^2$ leading to the Grashof number $Gr$ is not justifiable by physical principles, nor is the value $Gr=4\times10^7$ for the $\II\rightarrow\TT$ transition. In addition, although their $\sin\theta Re^2$ transition scaling agreed well with their data, it does not appear to agree with our more recent and comprehensive data  obtained in a smaller duct (figure~\ref{fig:regime-data-input-params}). We believe that the above points motivate the need for a revised scaling based on sound physical principles that are verified experimentally.

The qualitative classification into four discrete regimes introduced by \cite{macagno_interfacial_1961} and ML14 is an important first step in the study of the dynamics of sustained stratified shear flows. The presence or absence of interfacial waves, of small-scale structures indicative of turbulence, of spatio-temporal intermittency can all easily be picked by the eye using simple shadowgraph visualisation or dye visualisation (PLIF) and provide valuable `order one' information about the asymptotic (long-term, i.e. over hundreds of ATU) behaviour of the underlying dynamical system. 
Our novel volumetric 3D-3C measurements now allow us to complement these  \emph{qualitative} observations with \emph{quantitative} analyses of flows in each regime to investigate in more details their steady-state (asymptotic) dynamical equilibria.


We thus reformulate the aim of this paper introduced in \S~\ref{sec:intro} more specifically as:
finding a quantitative, physical basis explaining the different qualitative asymptotic behaviours of such sustained stratified shear flows (i.e. the `flow regimes'). Analysis of the past literature and our experimental observations suggest that the two leading non-dimensional input parameters of interest are $\theta$ and $Re$ ($A$ and $Sc$ playing lesser roles), hence we shall focus on them exclusively and seek transition curves of the form $\reg(\theta,Re)=$~const.

To tackle this aim, the rest of the paper is organised as follows. In \S~\ref{sec:framework}, we derive from first principles a framework of energy budget analyses suited to our 3D-3C measurements. In \S~\ref{sec:exp-validation}, we compare predictions for regime transition based on this framework to our experimental data. In \S~\ref{sec:3d}, we further develop this framework and the analysis of experimental data to get a deeper understanding of the relation between flow regimes and energetics. Finally, we summarise our findings and suggest future directions in \S~\ref{sec:ccl}.

\section{The energetics framework} \label{sec:framework}

In this section we introduce the theoretical framework to analyse the energetics of SID flows. We start by deriving the time evolution equations for the kinetic energy and potential energy, first as local quantities in \S~\ref{sec:local-budgets}, and then averaged in a control volume in \S~\ref{sec:avg-budgets}. To jump to the result of this section, see equations \eqref{dKdt-final} and \eqref{dPdt-final} and figure~\ref{fig:energetics-1}. We then estimate the transfer terms between kinetic and potential energies and simplify the budgets in \S~\ref{sec:estimations}.  Finally, we focus on one particular simplified budget in order to formulate an hypothesis regarding the regime transitions in \S~\ref{sec:implications}.

\subsection{Local energy budgets} \label{sec:local-budgets}

The governing equations on which all subsequent analyses are based are the incompressible Navier-Stokes equation under the
Boussinesq  approximation coupled to the advection-diffusion of density. Under the notation and conventions adopted in \S~\ref{sec:setup}, they take the following non-dimensional form
\begin{subeqnarray}\label{eq-motion}
\bnabla \cdot \uu &=& 0, \slabel{eq-motion-1}\\ 
\p_t \uu + \uu \cdot \bnabla \uu &=& -\bnabla p +  Ri \, ( -\cos \theta \,\mathbf{\hat{z}}  + \sin \theta \, \mathbf{\hat{x}})\rho  
+ \frac{1}{Re} \bnabla^2 \uu, \slabel{eq-motion-2}\\
\p_t \rho + \uu \cdot \bnabla \rho &=& \frac{1}{Re \, Sc} \bnabla^2 \rho. \slabel{eq-motion-3}
\end{subeqnarray}
where we recall that $Ri=1/4$ and $Sc=700$.

\subsubsection{Kinetic energy}

We first consider the kinetic energy field $\mathcal{K}$, defined as
\begin{equation} \label{definition_K}
\mathcal{K}(\xx,t) \equiv \frac{1}{2} u_i u_i,
\end{equation}
where, here and in the following, we adopt the summation convention over repeated indices. The evolution of  $\mathcal{K}$ is obtained by the dot product of the momentum equation \eqref{eq-motion-2} with $\uu$. Using incompressibility \eqref{eq-motion-1} and standard manipulations, we obtain
\begin{equation} \label{kinetic-energy-field-evolution}
\frac{\p \mathcal{K}}{\p t} = \phi_\mathcal{K}^\textrm{adv} +\phi_\mathcal{K}^\textrm{pre} +\phi_\mathcal{K}^\textrm{vis} +  \mathcal{B}_x - \mathcal{B}_z - \epsilon,
\end{equation}
where the boundary fluxes due to advection $\phi_\mathcal{K}^\textrm{adv}$, pressure work $\phi_\mathcal{K}^\textrm{pre}$, viscous work $\phi_\mathcal{K}^\textrm{vis}$ are
\begin{equation}\label{K-boundary-fluxes}
\phi_\mathcal{K}^\textrm{adv}  \equiv   \frac{\p}{\p x_i}(-u_i \mathcal{K}) , \qquad
\phi_\mathcal{K}^\textrm{pre}   \equiv   \frac{\p}{\p x_i}(-u_i p) , \qquad
\phi_\mathcal{K}^\textrm{vis}  \equiv   \frac{2}{Re}  \frac{\p }{\p x_j} (u_i \s_{ij}),
\end{equation}
and where the volumetric horizontal buoyancy fluxes $\mathcal{B}_x$, vertical buoyancy flux $\mathcal{B}_z$ and viscous dissipation $\epsilon$ are
\begin{equation}\label{K-volume-fluxes}
\mathcal{B}_x  \equiv  Ri \, \sin \theta \, \rho u, \qquad
\mathcal{B}_z   \equiv  Ri \,  \cos \theta \,  \rho w , \qquad
\epsilon \equiv  \frac{2}{Re} \s_{ij} \s_{ij}.
\end{equation}
The symmetric strain rate tensor is $\s_{ij} \equiv (\p_{x_i}u_j + \p_{x_j}u_i)/2$, and the dissipation rate is positive definite $\epsilon > 0$ .

\subsubsection{Potential energy}

Next, we consider the potential energy field $\mathcal{P}$, defined as
\begin{equation} \label{definition_P}
\mathcal{P}(\xx,t) \equiv Ri \, (z \cos \theta - x \sin \theta) \rho,
\end{equation}
since the duct $(x,y,z)$ coordinate system is tilted at angle $\theta$ with respect to the direction of gravity.
The evolution of $\mathcal{P}$ is obtained by standard manipulations of the density conservation equation \eqref{eq-motion-3} as
\begin{equation}\label{potential-energy-field-evolution}
\frac{\p \mathcal{P}}{\p t} \ = \phi_\mathcal{P}^\textrm{adv} +\phi_\mathcal{P}^\textrm{dif} +\phi_\mathcal{P}^\textrm{int} - \mathcal{B}_x + \mathcal{B}_z,
\end{equation}
where we recover the buoyancy fluxes $\mathcal{B}_x, \, \mathcal{B}_z$ defined in \eqref{K-volume-fluxes}, and where the boundary fluxes of $\mathcal{P}$ due to advection $\phi_\mathcal{P}^\textrm{adv}$, diffusion $\phi_\mathcal{P}^\textrm{dif}$, and conversion of internal energy $\phi_\mathcal{P}^\textrm{int}$ are
\begin{eqnarray}\label{P-boundary-fluxes}
\phi_\mathcal{P}^\textrm{adv}  &\equiv&   \frac{\p}{\p x_i}(-u_i \mathcal{P}), \nonumber \\
\phi_\mathcal{P}^\textrm{dif}   &\equiv&    \frac{Ri}{Re \, Sc} \frac{\p }{\p x_i} \Big\{(z \cos \theta - x \sin \theta) \frac{\p \rho}{\p x_i}\Big\},  \\
 \phi_\mathcal{P}^\textrm{int}  &\equiv&   \frac{Ri}{Re \, Sc} \Big\{ \cos \theta \frac{\p \rho}{\p z} - \sin \theta \frac{\p \rho}{\p x} \Big\}. \nonumber 
\end{eqnarray}

\subsection{Volume-averaged energy budgets} \label{sec:avg-budgets}

We now consider the control volume $V$, a rectangular parallelepiped bounded by the four duct cross-sectional walls at $y,z=\pm 1$ of arbitrary non-dimensional length $\ell \in [0, 2A]$ centred around $\bar{x}$, i.e. $V=(x,y,z) \in [ \bar{x}-\ell/2,\bar{x}+\ell/2 ] \times [-1,1] \times [-1,1]$ ($V$ has a volume equal to $\ell \times 2 \times 2 = 4\ell$). When applied to our 3D-3C data, the control volume $V$ will be the measurement volume shown in figure~\ref{fig:setup}.

\subsubsection{Kinetic energy}

We define the volume-averaged kinetic energy $K$ as
\begin{equation} \label{K-def}
K(t) \equiv  \langle \mathcal{K} \rangle_{x,y,z} \equiv \frac{1}{4\ell}\int_V \mathcal{K}  \, \d V = \frac{1}{4\ell} \int_{-1}^{1}\int_{-1}^{1} \int_{\bar{x}-\ell/2}^{\bar{x}+\ell/2} \mathcal{K}  \, \d x \, \d y \, \d z,
\end{equation}
where, here and henceforth, $\langle \cdot \rangle_{x,y,z}$ denotes averaging over the control volume $V$.

We obtain the evolution equation of $K$ by volume-averaging \eqref{kinetic-energy-field-evolution}. The volume-averaged boundary fluxes $\langle \Phi^\textrm{adv}_\mathcal{K}\rangle_{x,y,z}$, $\langle \Phi^\textrm{pre}_\mathcal{K}\rangle_{x,y,z}$, $\langle \Phi^\textrm{vis}_\mathcal{K} \rangle_{x,y,z}$ are simplified by the divergence theorem and the use of the no-slip boundary conditions $u_i =0$ on the four solid duct boundaries $y,z=\pm 1$. All mean gradients along $y$ and $z$ therefore cancel, and the mean gradients along $x$ take the general form $(1/\ell)\langle \cdot \rangle_{y,z}|_{L-R}$, where $\cdot |_{L-R}$ denotes the difference between the value of $\cdot$ on the left boundary of the volume (`L', $x=\bar{x}-\ell/2$) and its value on right boundary of the volume (`R', $x=\bar{x}+\ell/2$). We are left with
\begin{equation}  \label{dKdt-final}
\frac{dK }{dt}  =  \Phi_K^{\mathrm{adv}} + \Phi_K^{\mathrm{pre}} +\Phi_K^{\mathrm{vis}}+ B_x - B_z - D,
\end{equation}
where the boundary fluxes of $K$, the volume-averaged buoyancy fluxes and  dissipation are respectively 
\begin{equation}
\begin{gathered}  \label{K_fluxes}
\Phi_K^{\mathrm{adv}}  \equiv \frac{1}{\ell} \langle u \mathcal{K} \rangle_{y,z} |_{L-R}, \qquad
\Phi_K^{\mathrm{pre}}  \equiv \frac{1}{\ell}\langle u p\rangle_{y,z} |_{L-R}, \qquad
\Phi_K^{\mathrm{vis}}  \equiv  - \frac{1}{\ell}\frac{2}{Re}  \langle u_i s_{i1} \rangle_{y,z} |_{L-R},  \\
B_x  \equiv \langle \mathcal{B}_x \rangle_{x,y,z}, \qquad
B_z  \equiv  \langle \mathcal{B}_z \rangle_{x,y,z},\qquad
D  \equiv  \langle \epsilon \rangle_{x,y,z}. 
\end{gathered}
\end{equation}

\subsubsection{Potential energy}

We define the volume-averaged potential energy $P$ by analogy with $K$ as
\begin{equation} \label{P-def}
P(t) \equiv  \langle \mathcal{P} \rangle_{x,y,z} \equiv \frac{1}{4\ell}\int_V \mathcal{P}  \, \d V = \frac{1}{4\ell} \int_{-1}^{1}\int_{-1}^{1} \int_{\bar{x}-\ell/2}^{\bar{x}+\ell/2} \mathcal{P}  \, \d x \, \d y \, \d z,
\end{equation}
By volume averaging \eqref{potential-energy-field-evolution} and using the no-slip boundary condition for velocity and no-flux boundary condition for density, we write the evolution of $P$ as
\begin{equation} \label{dPdt-final}
\frac{d P}{d t} = \Phi_P^{\mathrm{adv}} +\Phi_P^{\mathrm{dif}} + \Phi_P^{\mathrm{int}}  - B_x + B_z,
\end{equation}
where the boundary fluxes of $P$ are
\begin{subeqnarray}  \label{P_fluxes}
\Phi_P^{\mathrm{adv}}  &\equiv& Ri \frac{1}{\ell} \big( \cos \theta \,  \langle z \rho u  \rangle_{y,z} |_{L-R}  - \sin \theta \,   \langle x \rho u\rangle_{y,z} |_{L-R}  \big), \slabel{phi-P-adv} \\
\Phi_P^{\mathrm{dif}}  &\equiv&
 \frac{Ri}{Re \, Sc} \frac{1}{\ell}  \Big(     \sin \theta \,  \langle x \frac{\p \rho}{\p x} \rangle_{y,z}\big|_{L-R} - \cos \theta \, \langle z \frac{\p \rho}{\p x} \rangle_{y,z}\big|_{L-R} \Big), \slabel{phi-P-dif} \\  
\Phi_P^{\mathrm{int}}  &\equiv&  \frac{Ri}{Re \, Sc} \Big( -\frac{1}{\ell} \sin \theta \langle \rho \rangle_{y,z}|_{L-R} + \frac{1}{2} \cos \theta \langle \rho \rangle_{x,y}|_{B-T}  \Big), \slabel{phi-P-int}
\end{subeqnarray}
where by analogy with $ \cdot |_{L-R}$, we denote by $\cdot  |_{B-T}$ the difference between the value of $\cdot$ at the bottom (`B', $z=-1$) and at the top (`T', $z=1$).

\subsubsection{Summary and schematics}

\begin{figure}
\centering
\includegraphics[width=0.5\textwidth]{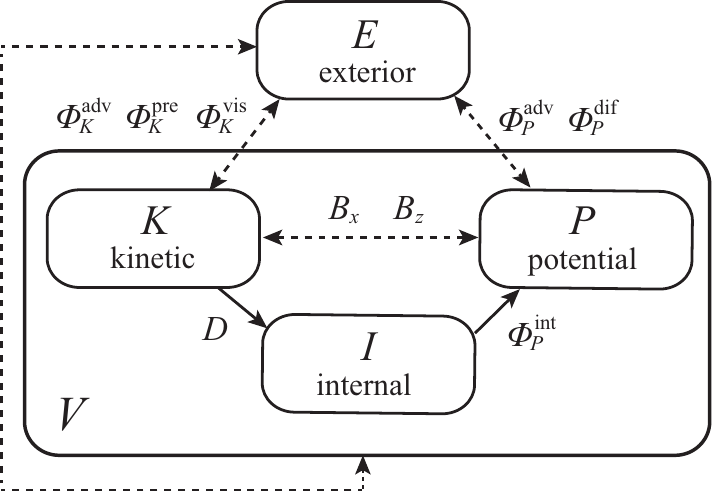}
\caption{Schematics of the \emph{a priori} complete energy budgets in a control volume $V$. The $V$-averaged kinetic $K(t)$, potential $P(t)$ and internal $I(t)$ energy reservoirs exchange energy with one another via internal fluxes and with the exterior $E$ via boundary fluxes. Solid arrows indicate irreversible (i.e. sign-definite) transfer, and dashed arrows indicate \emph{a priori} reversible (i.e. sign-indefinite) transfer, until proven otherwise later. The \emph{a priori} reversible transfer between $E$ and $I$ is acknowledged but was not explicitly derived in the text since it is not central to the discussion.}\label{fig:energetics-1}
\end{figure}

The evolution equations -- or `budgets' -- for the volume-averaged kinetic energy $K$ (see \eqref{dKdt-final} and \eqref{K_fluxes}) and potential energy $P$ (see  \eqref{dPdt-final} and \eqref{P_fluxes}) are summarised schematically in figure~\ref{fig:energetics-1}.

In addition to the kinetic energy $K$ and potential energy $P$ reservoirs, the fluid contained in the volume $V$ has an internal energy reservoir $I$ that we have hitherto not explicitly considered. As we shall see in \S~\ref{sec:simplified-budgets}, we do not need to do so since the evolution of $I$ is (to a very good approximation) slaved to that of $K$ and does not feed back on either $K$ or $P$.

These three reservoirs exchange energy via \emph{internal fluxes}: $K$ and $P$ exchange energy with one another via \emph{a priori} reversible (i.e. sign-indefinite) buoyancy fluxes $B_x$, $B_z$; $K$ is irreversibly dissipated at a positive-definite rate $D>0$ to $I$; and $I$ is irreversibly converted by molecular diffusion at a positive-definite rate $\Phi_P^{\textrm{int}}>0$ to $P$ (this conversion does not necessitate macroscopic fluid motions). 
In addition, $K$, $P$ and $I$ also exchange energy via a number of \emph{boundary fluxes} with the exterior (denoted by $E$). These boundary fluxes are all \emph{a priori} reversible (i.e. sign-indefinite).  (Note that the boundary flux of $I$ was not explicitly considered in the above discussion but we deduce its existence by the necessity to close the $I$ budget.)

The steady character of the sustained forcing in the SID experiment ensures that, \emph{when averaged over a sufficiently long time period, each energy reservoir must be in steady state}. In other words, the time-averaged budgets are `closed', in the sense that they all cancel:
\begin{equation} \label{steady-state}
\Big\langle \frac{dK}{dt} \Big\rangle_t \approx \Big\langle \frac{dP}{dt} \Big\rangle_t  \approx \Big\langle \frac{dI}{dt} \Big\rangle_t  \approx 0,
\end{equation}
where $\langle \cdot \rangle_t \equiv (1/\tau) \int_0^\tau \cdot \, \d t$ denotes averaging over the recorded data (or `duration of an experiment') $\tau$. We expect this steady state \eqref{steady-state} to be  a very good approximation, certainly over periods of $O(10^2-10^3$~ATU$)$ (the typical duration of an experiment), and presumably even over smaller periods of $O(10$~ATU$)$ in the relatively steady \textsf{L} and \textsf{H} regimes.

These budgets are related to other energetic analyses applied to numerical simulations in the literature (see e.g. \citealp[\S~4]{winters_available_1995}), but have a number of features that make them unique to SID experiments: \emph{(i)} the presence of a tilt angle $\theta>0^\circ$ introducing the crucial horizontal buoyancy flux $B_x$; \emph{(ii)} the presence of solid boundaries at $y,z=\pm 1$ cancelling the boundary fluxes along $y$ and $z$; \emph{(iii)} the absence of a periodic boundary condition in the $x$ direction introducing non-zero boundary fluxes along $x$ (contrary to most numerical simulations); and \emph{(iv)} the asymptotic steadiness of all reservoirs due to the sustained forcing discussed above.

In the remainder of the paper, we make the approximation that
\begin{equation} \label{small-angle-approx}
\cos \theta \approx 1 \quad \textrm{and} \quad  \sin \theta \approx \theta,
\end{equation}
which is accurate to better than $0.5~\%$ for the angles considered in this paper ($\theta \leq 6^\circ$). Unless explicitly specified, $\theta$ will now be expressed in radians.

\subsection{Estimations and simplified budgets} \label{sec:estimations}

In this section we give physical interpretation of each of the fluxes relevant to SID flows in order to determine their sign, relative magnitude, and eventually build a simplified picture of the time- and volume-averaged energetics of SID flows.

\subsubsection{The two-layer hydraulic model} \label{sec:two-layer-model}

\begin{figure}
\centering
\includegraphics[width=0.80\textwidth]{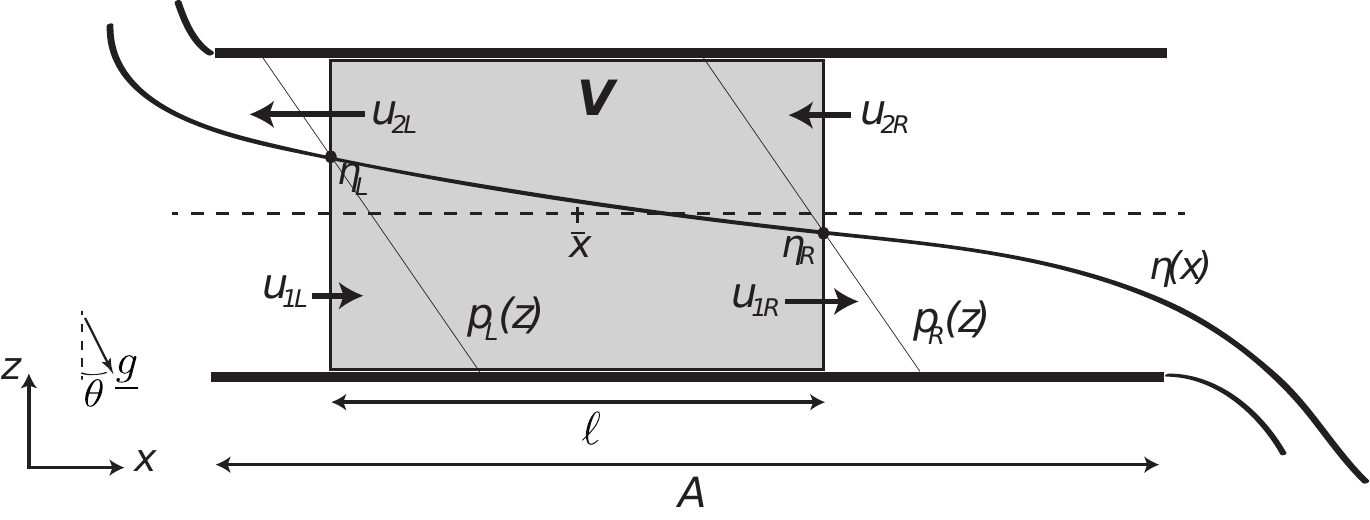}
\caption[Schematics and notation used for the evaluation of boundary fluxes]{Schematics and notation used for the evaluation of boundary fluxes under hydraulic assumptions. The control volume $V$, centred on $\bar{x}$ and of length $\ell$, is shaded in grey, and as before, 1 (resp. 2) denotes the lower (resp. upper) layer, and L (resp. R) denotes the left (resp. right) boundary of $V$. The interface has position $\eta(x)$ (solid curve) with respect to the neutral level $z=0$ (dashed). Note the hydrostatic pressure distributions $p_L(z)$ and $p_R(z)$ at the L and R boundaries (thin solid lines), with $p=0$ along the interface. }\label{fig:energetics-2}
\end{figure}

Consider the two-layer hydraulic model sketched in figure~\ref{fig:energetics-2}. The left (`L') boundary of the volume $V$ (shaded in grey) has a lower layer velocity $u_{1L}>0$, an upper layer velocity $u_{2L}<0$, and the right (`R') boundary of $V$ has a lower layer velocity $u_{1R}>0$, and an upper layer velocity $u_{2R}<0$. The position of the interface $\eta(x)$ (black solid curve) defined positive above the midplane $z=0$ (black dashed line) takes the respective values of $\eta_L$ and $\eta_R$ at each boundary. In agreement with hydraulic theory, and to make the following calculations easier, we further assume a steady streamwise velocity profile uniform in each layer (i.e. depending only on $x$), and a hydrostatic pressure distribution where the reference pressure is 0 all along the interface $p(x,z=\eta(x)) =0$ (after subtracting the hydrostatic streamwise pressure gradient due to $\theta \neq 0$). The local hydrostatic gradient is thus $\p_z p = Ri \, \rho = (1/4) \rho$ (where in the lower layer $\rho_1=1$, in the upper layer $\rho_2=-1$), giving a pressure distribution $p(x,z) = (1/4) \{ \eta(x)-z\}$ (shown as thin black solid lines). 

In order to gain insight into this model, consider its corresponding streamwise momentum equation (including viscous effects):
\begin{equation} \label{x-momentum}
4\uu \cdot \bnabla u = \underbrace{- \eta'(x)}_{\substack{\text{hydrostatic} \\ \text{forcing}}}+ \underbrace{ \theta \, \rho}_{\substack{\text{gravitational} \\ \text{forcing}}} + \frac{4}{Re} \bnabla^2 u,
\end{equation}
where $\rho(x,z) = \mathrm{sgn}(\eta(x) -z) = \pm 1$ by definition of $\eta(x)$. Since each layer convectively accelerates (and thus becomes thinner) in the direction in which is it flowing, the interface position $\eta$ must be a monotonically decreasing function of $x$: $\eta'(x)<0$ for all $x$. Since in addition $\eta \in [-1,1]$, the average slope on the scale of the whole duct (taking $\ell = 2A$) must be smaller than $2/2A = \alpha$, where we define the inverse aspect ratio of the duct as
\begin{equation} \label{definition-alpha}
\alpha \equiv A^{-1}
\end{equation}
We therefore have $\langle |\eta'(x)|\rangle_x < \alpha$, i.e. an upper bound  on the magnitude of the average slope and, therefore, on the magnitude of the horizontal pressure gradient in \eqref{x-momentum}. This bound holds for any sufficiently large volume $V$ not centred in the immediate vicinity of the ends of the duct  (where $|\eta'|$ may be large and the hydrostatic assumption may break down). 
Consequently, in such a control volume, a sufficient condition ensuring that the contribution of the \emph{gravitational} forcing in \eqref{x-momentum}  is always greater than the contribution of the \emph{hydrostatic} forcing is that the tilt angle $\theta$ is `large', which, in this paper, is understood as being large relative to the `geometrical' angle of the duct $\alpha$, i.e.
\begin{equation}
\theta > \alpha,
\end{equation}
For the duct discussed in this paper $\alpha = 1/30 \approx 2^\circ$. (Note that because of the length of the duct considered in this paper, a large tilt angle $\theta > 2^\circ$ is still compatible with our approximation \eqref{small-angle-approx}.)

A more accurate way to analyse the relative importance of the various terms in \eqref{x-momentum}, including the viscous friction in $\bnabla^2 u$, is through the framework of \emph{frictional two-layer hydraulic theory}. Originally proposed by \cite{schijf_theoretical_1953}, and later formalised by \cite{gu_frictional_2001,gu_analytical_2005}, this theory combines the hydraulic description of two-layer flows (see e.g. \cite{armi_hydraulics_1986}) with frictional stresses at solid boundaries and at the interface created by the inevitable $(y,z)$ dependence of the underlying velocity profiles. By parameterising the local loss of streamwise momentum due to these stresses by the local uniform model velocities $u_1(x),~u_2(x)$ using a small number of non-dimensional `friction' parameters, an expression for the local slope of the interface $\eta(x)$ can be derived. An adaptation of this theory to SID flow can be found in L18,~Chapter~5 but falls outside the scope of this paper. Here we limit ourselves to discussing the simple result that at the middle point of the duct ($x=0$) the interfacial slope is proportional to
%
\begin{equation}
\eta'(0) \propto \theta - F,
\end{equation}
where $F$ is the so-called `friction slope', a complicated expression combining wall and interfacial stress parameters. The above equation can be interpreted as follows: the viscous frictional stresses acting at the walls and at the interface parameterised in $F$ tend to make the interface slope \emph{downwards} (momentum sink), whereas the positive gravitational forcing $\theta>0$ tends to make the interface slope \emph{upwards} (momentum source). It follows that:
\begin{itemize}
\item When $0<\theta \ll F$, viscous friction in the duct makes the interface slope downwards, but as discussed above, with a magnitude that cannot exceed the duct geometrical slope: $F<\alpha$. The friction $F$ is largely independent of $\theta$, which does not play a significant dynamical role. We call such flows \emph{lazy} flows (figure~\ref{fig:lazy-forced}\emph{(a)}). 

\begin{figure}
\centering
\includegraphics[width=\textwidth]{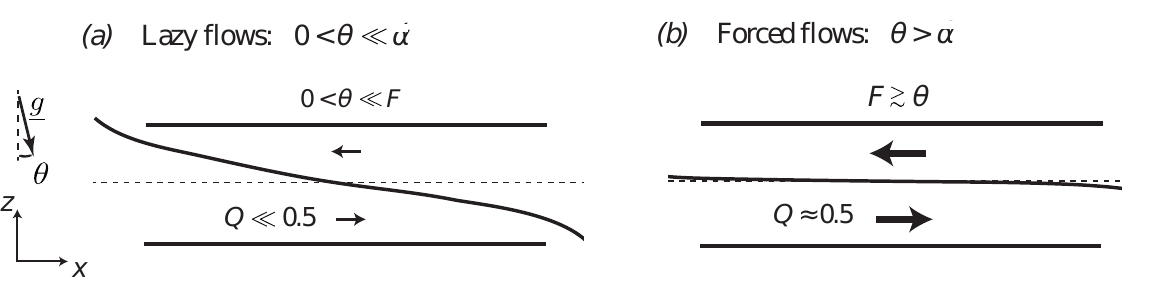}
    \caption{Qualitative distinction based on frictional hydraulic theory between \emph{(a)} `lazy' flows (at small tilt angles $\theta$), in which viscous effects in $F$ dominate over the gravitational forcing by $\theta$; and \emph{(b)} `forced' flows (at large tilt angles $\theta$) in which both effects are in balance, leading to a relatively flat interface throughout the duct and $Q\approx 0.5$.}\label{fig:lazy-forced}
\end{figure}

\item As $\theta$ is increased, the gravitational forcing makes the interface become increasingly horizontal (i.e. parallel to $x$) until it becomes nearly horizontal ($\eta'(0) \lesssim 0$) as $\theta$ approaches $F$ from below. As $\theta$ is further increased above this initial value of $F$, the friction $F$ \emph{must increase} to follow $\theta$ very closely to enforce the necessary condition that the interface slopes downwards. This qualitative change in the behaviour of the friction $F$, now directly dependent on $\theta$, occurs at the latest when $\theta > \alpha$ (since initially $F<\alpha$), yet generally for smaller $\theta$ (depending on the initial, unknown, value of $F$). In this situation, $F \gtrsim \theta$ and the interface is relatively flat throughout the duct ($\eta'(x)\lesssim 0$ for all $x$). We call such flows \emph{forced} flows (figure~\ref{fig:lazy-forced}\emph{(b)}). 
\end{itemize}

We believe that our distinction between lazy and forced flows is an important modelling result for the study of two-layer exchange flows forced by a positive angle $\theta>0$. In the next section, we build on this distinction to derive a much-simplified budget.

\subsubsection{Simplified budgets}\label{sec:simplified-budgets}

Based on the simplified two-layer hydraulic model introduced above, we derived estimations of each term of the full energy budget (\eqref{K_fluxes},~\eqref{P_fluxes}) in appendix~\ref{sec:estimation-fluxes}.

\begin{figure}
\centering
\includegraphics[width=0.77\textwidth]{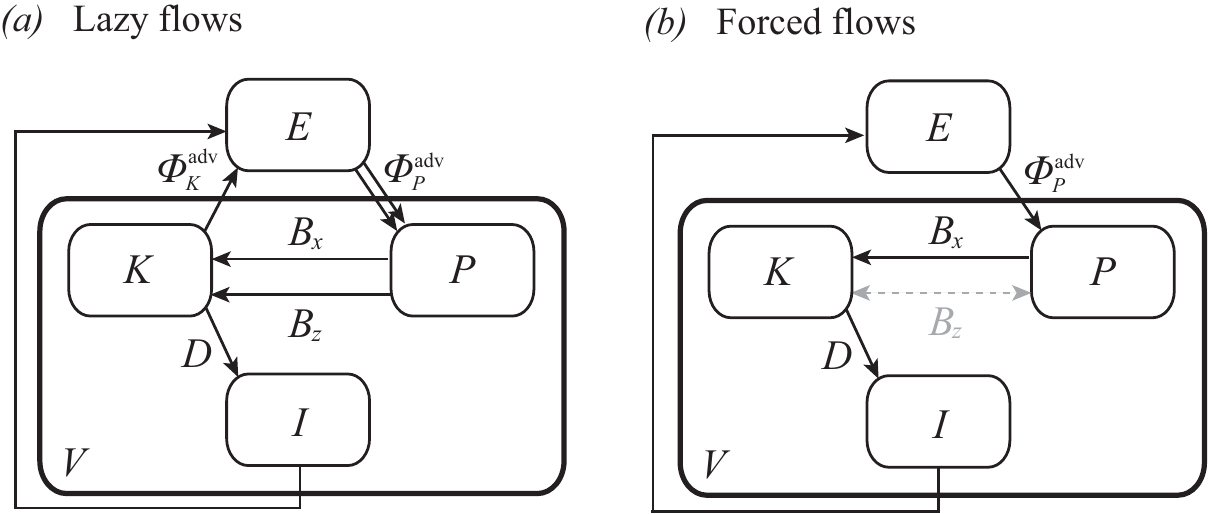}
\caption{Schematics of two simplified energy budgets. The energy fluxes in the general budget of figure~\ref{fig:energetics-1} were estimated in appendix~\ref{sec:estimation-fluxes} using the two-layer hydraulic model of figure~\ref{fig:energetics-2}, and led to two levels of simplifications for \emph{(a)} lazy flows and \emph{(b)} forced flows.}\label{fig:energetics-3}
\end{figure}

A \emph{first level of simplification} of the full budget presented in figure~\ref{fig:energetics-1} consists in neglecting the boundary fluxes $\Phi^\textrm{pre}_K$, $\Phi^\textrm{vis}_K$,  $\Phi^\textrm{dif}_P$, and $\Phi^\textrm{int}_P$ for the $Re$ and $Sc$ considered in this paper (as argued in appendix~\ref{sec:estimation-fluxes}). The resulting simplified budget for general SID flows, i.e. for lazy flows, is sketched in figure~\ref{fig:energetics-3}\emph{(a)}. In lazy flows (figure~\ref{fig:energetics-3}\emph{(a)}), all the energy in $V$ is supplied by the positive advective flux of $P$ ($\Phi^\textrm{adv}_P>0$) composed of hydrostatic and gravitational contributions (represented by a double arrow). This energy is transferred to $K$ by the horizontal buoyancy flux ($B_x>0$), equal to the gravitational contribution of $\Phi^\textrm{adv}_P$. We  previously argued that the vertical buoyancy flux $B_z$ was, in general, sign-indefinite, depending on the level of vertical motions in the flow. However it now becomes clear that, in order to close the budgets of lazy flows over sufficiently long times, $B_z$ must be a sink to $P$ and a source to $K$ ($B_z<0$), and it must equal the hydrostatic contribution of $\Phi^\textrm{adv}_P$ in magnitude. To balance these two distinct sources, $K$ has two distinct sinks: the advective flux $\Phi^\textrm{adv}_K<0$, and the viscous dissipation $-D<0$. (The internal energy reservoir $I$ has an energy source $D>0$, which in steady state, is balanced by a negative advective boundary flux to $E$.)

A \emph{second level of simplification} is possible in the special case of forced flows, as sketched in figure~\ref{fig:energetics-3}\emph{(b)}. We show in appendix~\ref{sec:estimation-fluxes} that in a `periodic' volume $V$ (expected when $\theta>\alpha$) the hydrostatic contribution of the source term $\Phi^\textrm{adv}_P$ and the advective flux $\Phi^\textrm{adv}_K$ both cancel. The budget becomes very simple: to a good approximation, the main source of $P$ is $\Phi^\textrm{adv}_P=(Q_m/4)\theta$, which corresponds exactly to its main sink (and therefore the main source of $K$) $B_x=\Phi^\textrm{adv}_P=(Q_m/4)\theta$. Therefore, although $B_z$ is truly sign-indefinite in this case and may be responsible for unsteady reversible energy transfers on short time scales, its \emph{temporal average} must cancel and become irrelevant in steady state over the duration of an experiment (hence we represent it by a grey dashed arrow). We thus conclude that, in steady state, $P$, $K$ (and $I$) all have only a single source and a single sink, which must all be equal in magnitude:
\begin{equation} \label{steady-state-dissipation}
\langle \Phi^\textrm{adv}_P \rangle_t = \langle B_x \rangle_t = \langle D \rangle_t = \frac{1}{4}\langle Q_m\rangle_t \theta.\end{equation}
This is one of the main modelling results of this paper. It states that the time- and volume-averaged energetics of forced flows in any control volume of the SID is reducible to a single flux which depends only on the magnitude of the mass flux exchanged between the two reservoirs $\langle Q_m\rangle_t$, and the tilt angle of the duct $\theta$. 

Another very attractive feature of forced flows is that the energy budgets we derived are valid in \emph{any} control volume $V$ in the duct regardless of its location $\bar{x}$ and length $\ell$. This is true as long as $V$ is not located in the immediate vicinity of the ends of the duct ($x=\pm A$) where the hydrostatic approximation is questionable and is sufficiently long (say $\ell \gg 1$) for the volume-averaging to make sense. Thus, by virtue of the $x$-periodicity of forced flows, the volume-averaged energetics of the whole duct are equal to that of any of its sub-volume and, in particular, of any sensible 3D-3C measurement volume.

\subsection{Implications: hypothesis for regime transitions} \label{sec:implications}

We now propose that the volume-averaged square norm of the (non-dimensional) strain rate tensor $S$, defined as
\begin{equation} \label{def-S}
S \equiv \langle \s_{ij}\s_{ij} \rangle_{x,y,z} = \frac{Re}{2} D,
\end{equation}
is a good candidate for a quantitative proxy of the flow regimes (as opposed to the viscous dissipation $D$ because of its $Re/2$ factor). In the remainder of the paper, we primarily focus on $S$ and refer to it as `viscous dissipation' for simplicity (which is the correct standard terminology with respect to the rescaled time coordinate $t^*\equiv t/(Re/2)$). Since the hydraulic controls at both ends of the duct limit the mean value of streamwise motions to $|u|_{x,y,z} = Q \lesssim 0.5$ and vertical motion must realistically be even smaller, we expect the range of spatial scales over which the strain rates act in $V$  to be the main variable of adjustment between flow regimes. We thus expect laminar flows with gradients over lengths of $O(1)$ to have $S = O(1)$ and increasingly turbulent flows with increasingly small-scale motions to have much larger gradients and $S \gg 1$.

It therefore appears natural to propose that the \LL, \HH, \II, \TT \, regimes correspond to increasingly large values of the time-averaged dissipation $\langle S \rangle_t$. This intuitive idea can be formalised using the regime function (see \eqref{def-reg}) as the following simple hypothesis:
\begin{equation} \label{reg-s2}
\reg = \reg(\langle S \rangle_{t}) ,
\end{equation}
where $\reg$ is a monotonically increasing function of $\langle S \rangle_{t}$ only. This hypothesis is general and does not assume that the flow is lazy or forced.

Our main modelling result \eqref{steady-state-dissipation} that the time- and volume-averaged dissipation $\langle S\rangle_t$ \emph{in forced flows} can be predicted from the knowledge of $\theta,~Re$ (input parameters) and $Q_m$ (output parameter) can be rewritten as
\begin{equation} \label{s2-Resintheta}
\langle S \rangle_t  = \frac{Re}{2} \langle D \rangle_t = \frac{1}{8}\langle Q_m\rangle_t \, \theta Re,
\end{equation}
Despite $Q_m$ being an output parameter, frictional hydraulic theory and extensive empirical evidence (see ML14, L18~\S~3.6 and figure~\ref{fig:regime-mSID-resintheta} below) suggest that the hydraulic limit of $Q_m \approx 0.5$ is usually a good approximation \emph{in forced flows}, so long as they are not excessively turbulent, since excessive turbulence and mixing acts to reduce $Q_m$ for very high values of  $\theta$ and $Re$ (as will be shown in figure~\ref{fig:regime-mSID-resintheta} below).

Therefore, the corollary of hypothesis \eqref{reg-s2} in the special case of forced flows is that regime transitions follow the simple scaling
\begin{equation} \label{s2-Resintheta-simplified}
\langle S \rangle_t  \approx \frac{1}{16} \, \theta Re,
\end{equation}
and \eqref{reg-s2} can be recast in terms of input parameters only
\begin{equation} \label{reg-thetaRe}
\reg = \reg(\theta Re) , 
\end{equation}
where $\reg$ is a monotonically increasing function of $\theta Re$ only. 

In the next section, we discuss experimental data  to examine the hypothesis \eqref{reg-s2} and its corollary  \eqref{reg-thetaRe}.

\section{Experimental validation} \label{sec:exp-validation}

In this section, we examine whether or not regime transitions:

\begin{itemize}
\item indeed scale with the non-dimensional group of parameters $\theta Re$ (the forced flow corollary of our physical hypothesis) using our  regime data in \S~\ref{sec:observed-scaling}; 

\item are indeed caused by increasing values of the time- and volume-averaged dissipation $\langle S \rangle_{t}$ (our underlying physical hypothesis) using our 3D-3C data in \S~\ref{sec:time-vol-avg-data}-\ref{sec:limitations}.
\end{itemize}
%

\subsection{Observed regime transitions scaling} \label{sec:observed-scaling}

\begin{figure}
\centering
\includegraphics[width=0.9\textwidth]{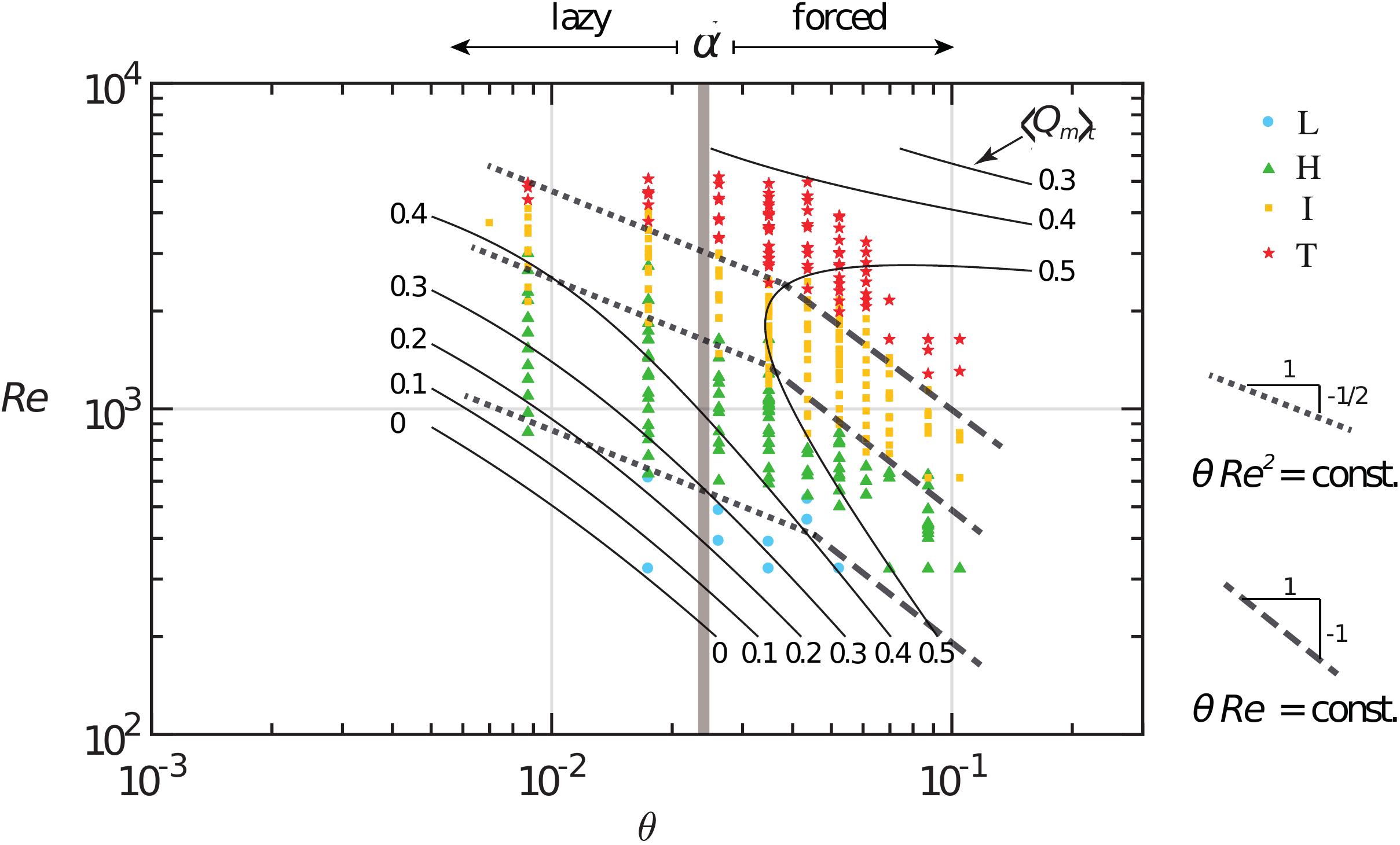}
    \caption{Scaling of regime transitions. The colour symbols are identical to figure~\ref{fig:regime-data-input-params}, and are plotted in the same $\theta-Re$ plane, but with $\theta$ in radians (also note the log-log scale, restricting us to $\theta>0^\circ$). The families of thick dotted and dashed lines represent approximate regime transition lines with respective scalings $\theta\,Re^2=$~const. and  $\theta Re=$~const. The vertical grey shading at $\theta=\alpha$ is the boundary between lazy and forced flows. The thin solid black contours: quadratic form fitting of 161 mass flux measurements of $\langle Q_m\rangle_t$. Six contours are shown in the range $0-0.5$ and they have been continued beyond the range covered by the data points used (note that no $0.6$ contour exists here). } \label{fig:regime-mSID-resintheta}
\end{figure}

To compare the scaling of the transitions in our experimental data with the model and predictions of the previous sections,  we plot in figure~\ref{fig:regime-mSID-resintheta} four distinct types of data in the $\theta-Re$ plane:
\begin{itemize}
\item The flow regime data of figure~\ref{fig:regime-data-input-params} using the same  symbols  (note that $\theta$ is expressed in radians here using a log scale, restricting us to $\theta>0$ data), 

\item Two families of thick lines indicating two distinct scaling: the dotted lines have slope $-1/2$ and indicate a power law scaling of the form $\theta Re^2=$~const. while the dashed lines have slope $-1$  and indicate a power law scaling of the form $\theta Re =$~const. These were set manually in order to best fit the data.

\item A vertical grey shading at $\theta =\alpha$ representing the upper bound for the expected boundary between lazy flows and forced flows (see \S~\ref{sec:two-layer-model}).

\item Thin black contours showing a fit of $\langle Q_m \rangle_t$ based on 161 mass flux measurements (see \S~\ref{sec:MF}). These data were then fitted by least-squares assuming a quadratic form in the $(\log \theta,\,\log Re)$ plane.
\end{itemize}

We make the following observations:

\begin{enumerate}
\item The mass flux data $\langle Q_m\rangle_t$ are best fitted by a quadratic form describing hyperbolas having a major axis of slope $-0.67$, i.e. an equation $\theta Re^{3/2}=$~const. This empirical scaling, and more generally, the function $\langle Q_m \rangle_t(\theta,Re)$, are not presently understood and fall outside the scope of the present study (see L18,~\S~3.6 for more details). Here, we limit ourselves to the empirical observations that:  \emph{(i)} for the `lazy' data ($\theta<\alpha$), as $\theta$ and $Re$ increases, $\langle Q_m \rangle_t$ increases from $\ll 0.5$ (\LL \ regime) to $\approx 0.5$ (\II \ and \TT \ regimes); \emph{(ii)} for the `forced' data ($\theta>\alpha$), $\langle Q_m\rangle_t \approx 0.5$. These two observations, given the fact that $Q_m \approx Q$ (except for the most turbulent data), are consistent with the theoretical predictions of \S~\ref{sec:two-layer-model}. 

\item In lazy flows, the regime data follow a $\reg \sim \theta \,Re^2$ scaling (dotted lines). The $\LL \rightarrow\HH$, $\HH \rightarrow\II$, and $\II\rightarrow\TT$  transitions curves are respectively \mbox{$\theta \,Re^2 =6\times10^3, \, 6\times 10^4, \, 2\times 10^5$}. This empirical `lazy flow scaling' is not consistent with the theoretical `forced flow scaling' predicted by the corollary \eqref{reg-thetaRe}, which is not surprising given the different energetics of lazy flows. This $\theta Re^2$ scaling is however consistent with the scaling proposed by ML14 (see \S~\ref{sec:ML14} and \eqref{grashof}), but this may be a coincidence that is not presently understood.

\item In forced flows, the regime data  follow a $\reg \sim \theta \,Re$ scaling (dashed lines).  The $\LL \rightarrow\HH$, $\HH \rightarrow\II$, and $\II \rightarrow\TT$  transitions are respectively \mbox{$\theta \,Re \approx 20, \, 50, \, 100$}. This empirical `forced flow scaling' is consistent with the corollary \eqref{reg-thetaRe} (and inconsistent with ML14).

\end{enumerate}

We have thus confirmed one of the features underlying the distinction between lazy and forced flows ($Q\approx Q_m < 0.5$ \emph{vs} $\approx 0.5$ respectively), as well as the regime transitions scaling in forced flows $\reg=\reg(\theta Re)$ (corollary \eqref{reg-thetaRe}), but showed that lazy flows followed a different (and still unexplained) scaling. 

In order to confirm the hypothesis \eqref{reg-s2} underlying the corollary, and thus to provide a physical basis for our understanding of regime transitions, we need to validate the energetics framework of \S~\ref{sec:framework}, and in particular, we need direct evidence that the energy budget of forced flows indeed follows the simplified model in figure~\ref{fig:energetics-3}\emph{(b)}. This is the subject of the next section.

\subsection{Experimental energy budgets} \label{sec:time-vol-avg-data}

We turn our attention to the energy budgets of 16 3D-3C experiments, whose input parameters, volume properties and resolution are detailed in table~\ref{tab:3d-3c-expts}. 
They include one experiment in the  $\LL$ regime ($\theta \,Re<20$, named `L1'), four in the $\HH$ regime ($20<\theta \,Re<50$,  `H1' to `H4'), eight in the $\II$ regime ($50<\theta \,Re<100$,  `I1' to `I8'), and three in the $\TT$ regime ($\theta \,Re>100$,  `T1'to `T3').

\begin{table}
  \begin{center}
\def~{\hphantom{0}}
\setlength{\tabcolsep}{7.5pt}
  \begin{tabular}{lcccccccccc}
      Name & $\theta$ (${}^\circ$)   &  $Re$  & $\theta Re$ &  \multicolumn{3}{c}{Volume properties} & \multicolumn{3}{c}{Resolution of data}  \\
             &   &   & & $\bar{x}$ & $\ell$ & $\tau$ & $\Delta x,\, \Delta z$ & $\Delta y$  & $\Delta t$   \\ [5pt]
L1 & 2 & 398 & 14 & $-12.2$ & 10.4 & 936 & 0.026 & 0.061 & 3.75 \\[5pt] 
H1 & 1 & 1455 & 25 & $-12.2$ & 10.4 & 459 & 0.025 & 0.053 & 2.29  \\ 
H2 & 5 & 402 & 35 & $-11.9$  & 10.8 & 302 & 0.025 & 0.074 & 1.03 \\ 
H3 & 2 & 1059 & 37 & $-12.4$ & 11.2 & 351 & 0.025 & 0.036 & 2.64  \\ 
H4 & 5 & 438 & 38 & $-12.0$ & 11.0 & 335 & 0.027 & 0.069 &  1.08  \\[5pt] 
I1 & 2 & 1466 & 51 & $-12.4$ & 11.2 & 508 & 0.026 & 0.036 &  3.65  \\ 
I2 & 2 & 1796 & 63 & $-12.4$ & 11.1 & 456 & 0.025 & 0.061 & 2.90 \\ 
I3 & 2 & 2024 & 71 & $-12.5$ & 11.1 & 722 & 0.025 & 0.063 & 3.28 \\ 
I4 & 6 & 777 & 81 & $-12.6$ & 7.73 & 248 & 0.019 & 0.057 & 1.65 \\ 
I5 & 5 & 956 & 83 &  $-11.0$ & 10.0 & 332 & 0.025 & 0.067 & 1.27 \\ 
I6 & 6 & 798 & 83 & $-12.6$ & 7.67 & 116 & 0.019 & 0.059 & 0.85 \\ 
I7 & 3 & 1580 & 83 & $-14.0$  & 7.49 & 223 & 0.018 & 0.056 & 1.68 \\ 
I8 & 5 & 970 & 84 & $-11.9$ & 11.8 & 250 & 0.026 & 0.054 & 1.69 \\[5pt] 
T1 & 3 & 2331 & 122 & $-14.0$ & 7.50 & 407 & 0.019 & 0.057 & 2.70  \\ 
T2 & 6 & 1256 & 131 & $-12.5$ & 7.66 & 203 & 0.019 & 0.057 & 1.34   \\ 
T3 & 5 & 1516 & 132 & $-11.9$ & 11.1 & 554 & 0.025 & 0.053 & 2.39 \\ 
  \end{tabular}
  \caption{List of the 16 3D-3C experiments used, showing the input parameters $\theta$ and $Re$, volume properties and resolution of data. In the second column only, $\theta$ is expressed in ${}^\circ$. Experiments are sorted by increasing $\theta Re $.}
\label{tab:3d-3c-expts}
  \end{center}
\end{table}

\begin{figure}
\centering
\includegraphics[width=0.9\textwidth]{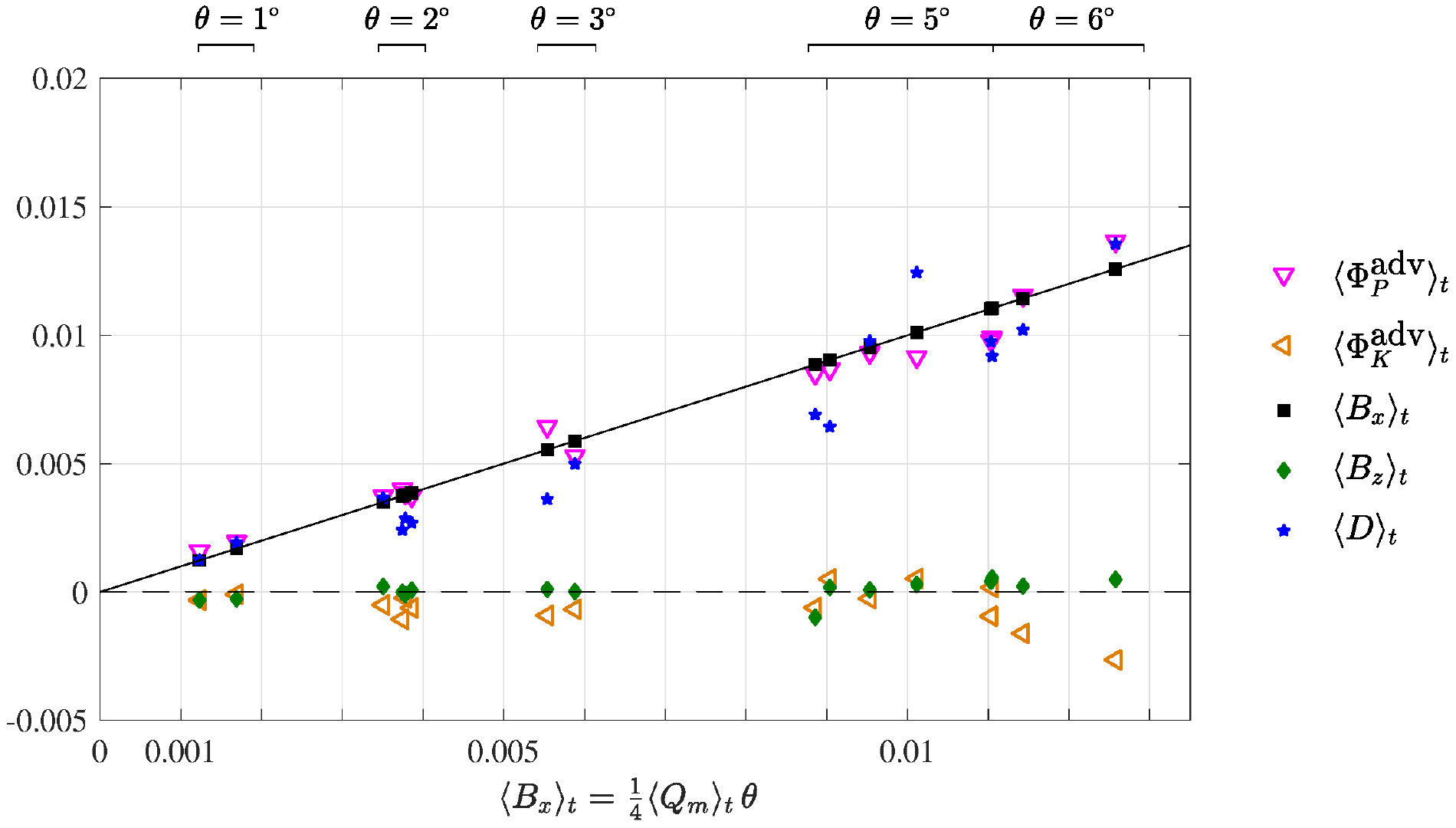}
    \caption{Experimental validation of the simple `forced flow' energetics model sketched in figure~\ref{fig:energetics-3}\emph{(b)}. Time-averaged energetics of the 16 3D-3C experiments in table~\ref{tab:3d-3c-expts}. Each flux retained in the general `lazy flow' model of figure~\ref{fig:energetics-3}\emph{(a)} is plotted against $\langle B_x \rangle_t$ (close to being the input parameter $\theta$), showing that, as expected for forced flows, $\langle \Phi^\textrm{adv}_P \rangle_t \approx \langle D\rangle_t \approx \langle B_x \rangle_t$ and $\langle B_z \rangle_t\approx \langle \Phi^\textrm{adv}_P \rangle_t\approx 0 $. } \label{fig:validation}
\end{figure}

In figure~\ref{fig:validation}, we plot the five main time-averaged energy fluxes of interest to validate the energetics model of \S~\ref{sec:framework} and figure~\ref{fig:energetics-3}: $\langle \Phi^\textrm{adv}_P\rangle_t$ (magenta triangles), $\langle \Phi^\textrm{adv}_K \rangle_t$ (orange triangles), $\langle B_x \rangle_t$ (black line and squares), $\langle B_z \rangle_t$ (green lozenges) and $\langle D \rangle_t$ (blue stars). In this plot, the vertical coordinate of each symbol represents the value of its respective flux, and its horizontal coordinate represents the value of the horizontal buoyancy flux $\langle B_x \rangle_t$ for this particular experiment. All fluxes are therefore effectively plotted against $\langle B_x \rangle_t$, whose definition $\langle B_x \rangle_t= (1/4)\langle Q_m\rangle_t \theta \approx \theta/8$ (assuming $Q_m\approx 0.5$) makes it closest to being an input parameter. Note that this choice of horizontal coordinate automatically groups the data by increasing values of $\theta$ (i.e. importantly \emph{not} by increasing $\theta\,Re$, thus not by regime). Note that the $\theta=2^\circ$ group of data includes a mix of $\LL$, $\HH$ and $\II$ flows,  the $\theta=5^\circ$ group includes $\HH$, $\II$ and $\TT$  flows and the $\theta=3^\circ$ and $\theta=6^\circ$ groups include $\II$ and $\TT$  flows.

We observe that $\langle \Phi^\textrm{adv}_P \rangle_t$ (main source of $P$) and $\langle D\rangle_t$ (main sink of $K$) closely follow the buoyancy flux $\langle B_x \rangle_t$ ($P\rightarrow K$ exchange) at all angles. The dissipation data show the greatest discrepancy (i.e. the blue stars lie further away from the black line and squares than the magenta triangles do) as we will explain in \S~\ref{sec:limitations}. 
We also verify that the advective flux of kinetic energy and the vertical buoyancy fluxes, which are only expected to be relevant in lazy flows, are indeed close to zero: $\langle \Phi^\textrm{adv}_K \rangle_t, \, \langle B_z\rangle_t \approx 0$ (see dashed line). 

In other words, the simplified budgets of figure~\ref{fig:energetics-3}\emph{(b)} for forced flows and our main prediction \eqref{steady-state-dissipation} that the energetics of SID flows are reducible to a single energy flux (that we may refer to as `power throughput') appear to be good approximations for $\theta \in [1^\circ, 6^\circ]$, that is, even when the necessary condition for forced flows $\theta>\alpha \approx 2^\circ$ does not hold.

\begin{figure}
\centering
\includegraphics[width=0.9\textwidth]{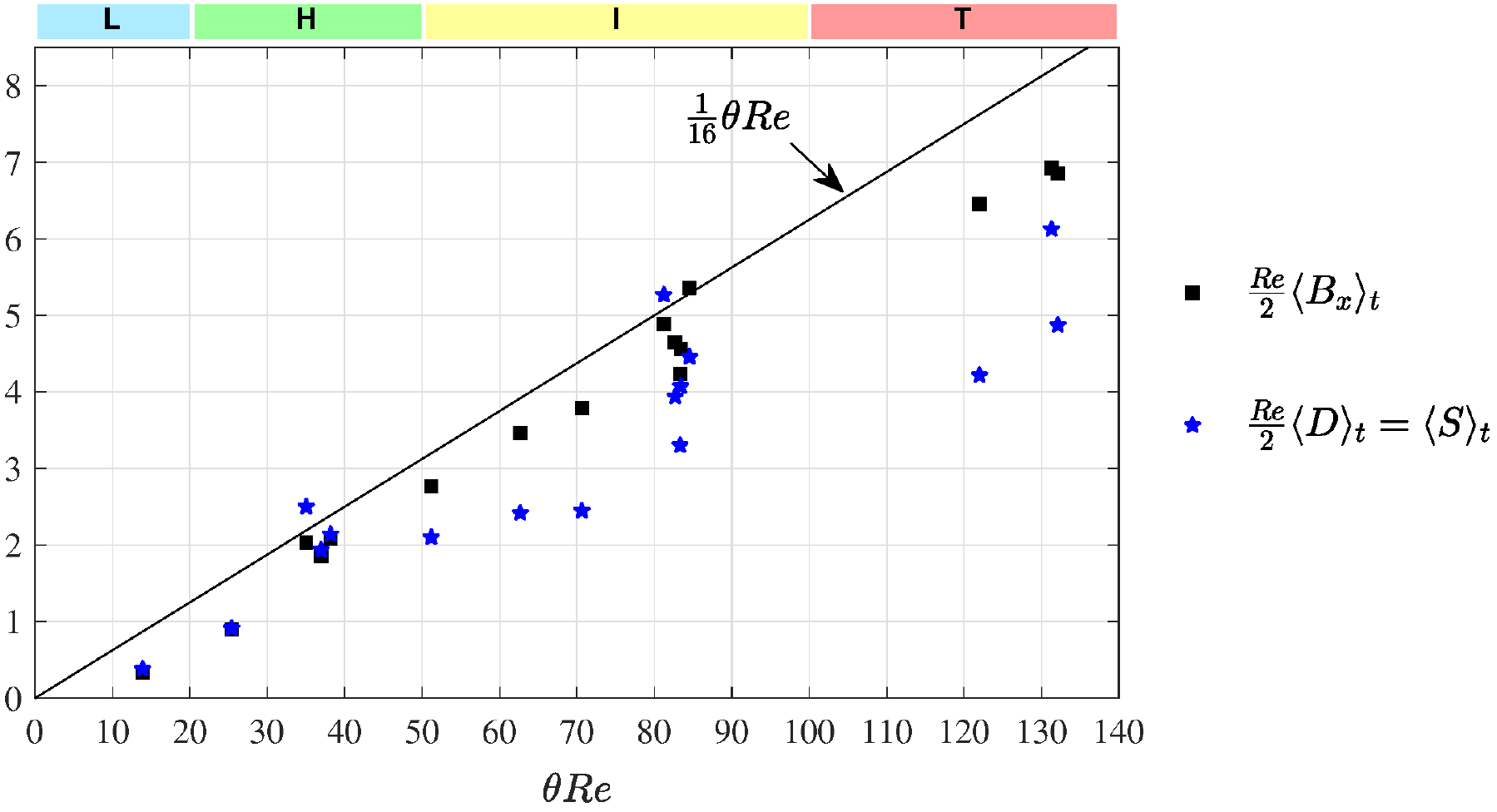}
    \caption{Dissipation and buoyancy flux data of figure~\ref{fig:validation} (same symbols) rescaled by $Re/2$ and plotted against the input parameter $\theta Re$ to test the corollary \eqref{s2-Resintheta-simplified} (black line). } \label{fig:dissipation}
\end{figure}

Although we do not show these results, we verified that the experimental time-averaged kinetic and potential energy budgets do indeed cancel to an excellent approximation: $\langle dP/dt \rangle_t \approx \langle dK/dt \rangle_t \approx0$ as hypothesised in \eqref{steady-state} (the flow has steady $P$ and $K$ reservoirs). However, it is clear from figure~\ref{fig:validation} that, for some experiments, these budgets do not cancel to such a good approximation when indirectly computed from the sum of experimentally-determined fluxes (i.e. $\langle dP/dt\rangle_t = \langle \Phi^\textrm{adv}_P\rangle_t -\langle B_x \rangle_t + \langle B_z\rangle_t$ and similarly $\langle dK/dt\rangle_t = \langle \Phi^\textrm{adv}_K\rangle_t +\langle B_x \rangle_t - \langle B_z\rangle_t -\langle D\rangle_t $ as per figure~\ref{fig:energetics-3}). This is due to the greater experimental errors in determining boundary fluxes and dissipation rates than in determining $dK/dt$ and $dP/dt$ directly.

In figure~\ref{fig:dissipation}, we re-plot the buoyancy flux and dissipation data of figure~\ref{fig:validation} (black squares and blue stars) rescaled by $Re/2$. The dissipation  $\langle S \rangle_t =(Re/2) \langle D \rangle_t$ is our hypothetical proxy for the flow regimes and we test its dependence on the transition parameter $\theta Re$ expected from the corollary \eqref{reg-thetaRe}. Plotted against this horizontal axis, the data are no longer grouped by angles (as was the case in figure~\ref{fig:validation}); rather they are grouped by increasing flow regimes (as shown by the coloured boxes at the top of the figure).

These data generally support the physical hypothesis that each flow regime corresponds to a  well defined range of $\langle S \rangle_t$ scaling with $\theta Re$. However the agreement with the simplified scaling  \eqref{s2-Resintheta-simplified} $\langle S\rangle_t \approx (1/16) \theta \,Re$ (black solid line) is not particularly impressive (blue stars lying below the black line in all but two experiments). This discrepancy has two causes: \emph{(i)} the approximation $\langle Q_m \rangle_t = 0.5$ is an upper bound for most experiments (black squares lying below the black line) as discussed in \S~\ref{sec:MF} and \S~\ref{sec:observed-scaling}; \emph{(ii)} the viscous dissipation is generally underestimated in experiments (blue stars lying below the black squares). We discuss the latter next.


\subsection{Experimental limitations in measuring the dissipation} \label{sec:limitations}

The previous section showed that, despite measurements showing that the kinetic energy reservoir was steady $(Re/2) \langle dK/dt\rangle_t \approx 0$, its sink $\langle S \rangle_t$ was generally measured to be smaller in magnitude than its source $(Re/2)\langle B_x \rangle_t$ in the $\II$ and $\TT$ regimes. This is due to at least three experimental limitations specific to measurements of the dissipation:

First, \emph{numerically}, the dissipation is the only flux that requires computing of flow field derivatives. Despite our use of a second-order accurate finite-difference scheme to compute the components of the strain rate tensor, experimental errors are bound to be amplified by derivations especially in the $\II$ and $\TT$ regimes where gradients are computed over small lengthscales;
    
\begin{figure}
\centering
\includegraphics[width=0.99\textwidth]{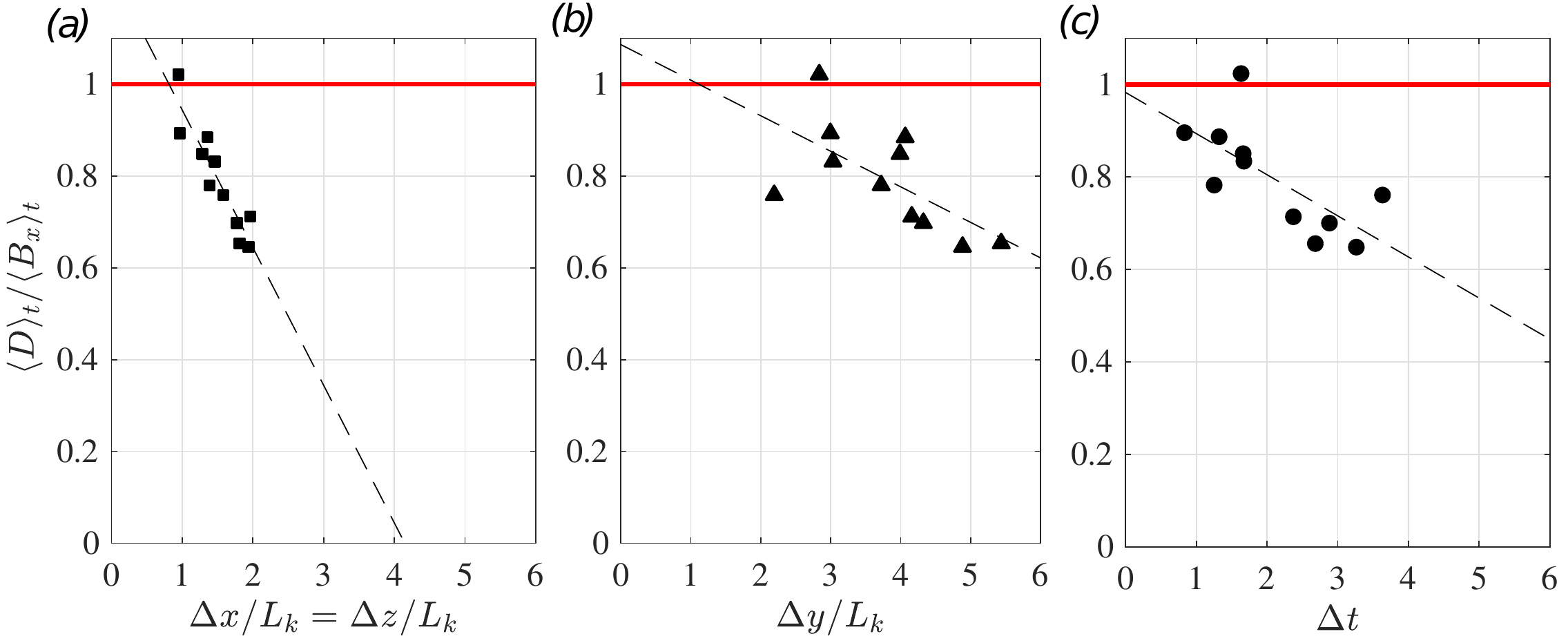}
    \caption{Effect of the spatio-temporal resolution of experiments on the accuracy of dissipation measurements in the $\II$ and $\TT$ experiments of table~\ref{tab:3d-3c-expts} and \mbox{figures~\ref{fig:validation}-\ref{fig:dissipation}}. Measurements converge towards the expected value ($\langle D\rangle_t = \langle B_x\rangle_t$, red line) for increased \emph{(a-b)} spatial resolution with respect to the Kolmogorov scale, and \emph{(c)} temporal  resolutions (better `freezing' of volumes). Dashed line represents the best linear fit.} \label{fig:resolution}
\end{figure}

Second, \emph{dynamically}, measurements of turbulent dissipation rates require a fine enough spatial resolution, i.e. a grid size $(\Delta x,\Delta y,\Delta z)$ small enough to capture the smallest dynamically active scale. It is generally acknowledged that the spectral content of dissipation becomes negligible below the Kolmogorov lengthscale, which is defined dimensionally as $L_k \equiv (\nu^3/\langle \epsilon \rangle_{x,y,z,t})^{1/4}$ (where, here and here only, $L_k$ and $\epsilon$ are dimensional). Because we know that the kinetic energy budget is closed, we use the estimated time- and volume-averaged dissipation of our corollary \eqref{s2-Resintheta-simplified} to estimate the non-dimensional Kolmogorov lengthscale as:
\begin{equation} \label{Lk_nondim}
L_k \equiv \frac{1}{(H/2)} \Bigg[ \frac{\nu^3}{2\nu \frac{g'H}{(H/2)^2}\langle\mathsf{s}_{ij}\mathsf{s}_{ij}\rangle_{x,y,z,t}} \Bigg]^{1/4} \approx 2^{3/4}  (Re^3 \sin \theta)^{-1/4}.
\end{equation}
For each of the 11 experiments in the $\II$ and $\TT$ regimes, we plot in figure~\ref{fig:resolution}\emph{(a-b)} the ratio $\langle B_x \rangle_t / \langle D\rangle_t$ against the spatial resolution normalised by the Kolmogorov lengthscale \eqref{Lk_nondim}: $\Delta x/L_k=\Delta z/L_k$ in panel $\emph{(a)}$ and  $\Delta y /L_k$ in panel $\emph{(b)}$.
We observe that the estimates of dissipation become more accurate (converging to the red horizontal line) as the spatial resolution approaches the Kolmogorov lengthscale (the dashed line is the best linear fit to the data and intercepts the red line at $\Delta x, \Delta z \approx L_k$). In other words, experiments featuring the largest discrepancy in figures~\ref{fig:dissipation} were the ones in which the spatial resolution of experimental measurements was not sufficient given the level of turbulence expected for their value of $\theta Re$. We note that this trend was not observed when the data were plotted against  $\Delta x, \Delta y, \Delta z$ alone (i.e. the Kolmogorov scale is important). This latter observation suggests that the lack of spatial resolution dominates over the numerical inaccuracies discussed in the previous paragraph.

Third, accurate measurements of flow gradients require our 3D-3C volumetric measurements to be as instantaneous as possible. As discussed in \S~\ref{sec:3d-3c} and appendix~\ref{sec:appendix-exp}, our scanning technique sets a lower bound on the non-dimensional time resolution $\Delta t$ over which a volume is constructed. These non-instantenous measurements inevitably distort turbulent flow structures. Figure~\ref{fig:resolution}\emph{(c)} quantifies this impact and demonstrates that better temporal resolutions with respect to an ATU (more `frozen' volumes) result in better estimates of $\langle S \rangle_t$ (the fit intercepts the red line at $\Delta t \approx 0$, as expected). The reason why such distortions lead to under-estimations (as opposed to over-estimations) of velocity gradients is still poorly understood. 

\section{Regimes, dissipation and three-dimensionality}\label{sec:3d}

In the previous section, we validated experimentally our hypothesis that regime transitions correlate with an increase in the non-dimensional, volume-averaged strain rate $S$ (that we refer to as `dissipation') and our corollary that they both scale with $\theta Re$. 

In this section, we seek to gain more insight 
investigating the link between flow energetics and three-dimensionality. We start by analysing the energy budgets of forced flows in more detail by subdividing the kinetic energy into a two-dimensional and a three-dimensional part in \S~\ref{sec:2d-3d-budgets}, before sketching them and discussing their implications for regime transitions in \S~\ref{sec:2d-3d-disc}. We then validate this framework using experimental data in \S~\ref{sec:2d-3d-experimental-data} and focus on spatial structures in \S~\ref{sec:2d-3d-structure}.

\subsection{Two-dimensional and three-dimensional kinetic energy budgets} \label{sec:2d-3d-budgets}

We start by defining, for any flow field $\phi$, a decomposition into a streamwise-averaged two-dimensional  component $\phi^{2d}$ and a complementary three-dimensional  component $\phi^{3d}$:
\begin{equation} \label{2d-3d-decomposition-1}
\phi(x,y,z,t) = \phi^{2d}(y,z,t) + \phi^{3d}(x,y,z,t) , 
\end{equation}
where
\begin{subeqnarray} \label{2d-3d-decomposition-2}
\phi^{2d}(y,z,t) &\equiv& \langle \phi \rangle_x, \\
\phi^{3d}(x,y,z,t)  &\equiv& \phi-\langle \phi \rangle_x.
\end{subeqnarray}
This decomposition is inspired from similar decompositions applied to direct numerical simulations (DNS) of stratified turbulence initiated by secondary instabilities developing on Kelvin-Helmholtz (KH) billows \citep{caulfield_anatomy_2000,peltier_mixing_2003,mashayek_zoo_2012,mashayek_zoo_2012-1,mashayek_time-dependent_2013,salehipour_turbulent_2015}. These studies typically decomposed the kinetic energy and associated fluxes into a one-dimensional part, corresponding to an initial base flow varying along $z$, a two-dimensional $(x,z)$ part corresponding to the primary KH instability, and a three-dimensional $(x,y,z)$ part corresponding to the `zoo' of secondary instabilities developing on the time-evolving KH billow. Our decomposition is slightly different in order to reflect the fact that, due to confinement by the duct boundaries, the SID `base flow' is an inherent two-dimensional function of $y$ and $z$ (for more details see LPZCDL18 \S~5.3).

Next, we define the volume-averaged 2D and 3D kinetic energies based on the respective velocity fields:
\begin{subeqnarray} \label{2d-3d-KE}
K^{2d}(t) &\equiv& \langle \mathcal{K}^{2d} \rangle_{y,z} \ \equiv  \frac{1}{2} \langle  u^{2d}_i u^{2d}_i \rangle_{y,z},  \\
K^{3d}(t) &\equiv& \langle \mathcal{K}^{3d} \rangle_{x,y,z} \equiv \frac{1}{2}  \langle u^{3d}_i u^{3d}_i \rangle_{x,y,z}.
\end{subeqnarray}
Importantly, we verify that the total kinetic energy is the sum of both components: $K = K^{2d}+K^{3d}$, since $\langle \mathcal{K} \rangle_x = \langle \mathcal{K}^{2d}\rangle_x + \langle \mathcal{K}^{3d}\rangle_x + u_i^{2d} \langle u_i^{3d} \rangle_x$ and $\langle u_i^{3d} \rangle_x=0$ by definition.
In order to write the evolution of $\mathcal{K}^{2d}$ and $K^{2d}$, we first $x$-average the momentum equation, which involves a number of gradients and divergence terms of the form
\begin{equation}
\Big \langle \frac{\p \phi }{\p x_i} \Big\rangle_x  =  \underbrace{\Big \langle \frac{\p \phi }{\p x} \Big\rangle_x }_{\textrm{mean gradient}}+  \frac{\p \langle \phi \rangle_x }{\p x_i}, 
\end{equation}
In this integration by parts, $\phi$ may represent $u_i u_j$ (convective term), $p$ (pressure gradient), or $u_i$ (diffusive term). At this point, the assumption of periodic boundaries in $x$, consistent with forced flows (see figure~\ref{fig:lazy-forced}\emph{(b)}), becomes essential in order to cancel all mean gradients along $x$ (the first term on the RHS) and make analytical progress (by avoiding very lengthy  expressions). 
Thus, \emph{under this essential periodic assumption}, we derive the following simple budgets:
\begin{subeqnarray}  \label{dK2d3ddt-final}
\frac{dK^{2d} }{dt^*} = \frac{Re}{2} \frac{dK^{2d} }{dt}   &=&  \frac{Re}{2}( B^{2d}_x - B^{2d}_z)  - S^{2d} - T,\\
\frac{dK^{3d} }{dt^*}  = \frac{Re}{2} \frac{dK^{3d} }{dt}  &=&   \frac{Re}{2}( B^{3d}_x - B^{3d}_z)  - S^{3d} + T,
\end{subeqnarray}
where the rescaled `fast' time $t^*\equiv t/(Re/2)$, previously introduced in \S~\ref{sec:implications}, is now used to facilitate general comparison between all experiments (making the horizontal buoyancy flux scale with $\theta Re$ and the rate of viscous dissipation be $S$ instead of $D$).
We define the above two-dimensional and three-dimensional buoyancy fluxes, dissipation, and the new transfer term $T$  between $K^{2d}$ and $K^{3d}$ as
\begin{subeqnarray}  \label{K2d3d_fluxes}
B_x^{2d}  &\equiv&  \frac{\theta}{4}    \langle \rho^{2d}u^{2d} \rangle_{y,z}, \qquad  B_z^{2d}  \equiv \frac{1}{4}   \langle \rho^{2d}w^{2d} \rangle_{y,z}, \qquad S^{2d}  \equiv    \langle \mathsf{s}^2_{2d} \rangle_{y,z}, \\
B_x^{3d}  &\equiv& \frac{\theta}{4} \langle \rho^{3d}u^{3d} \rangle_{x,y,z}, \ \quad B_z^{3d} \equiv \frac{1}{4} \langle \rho^{3d}w^{3d} \rangle_{x,y,z}, 
  \quad  \ S^{3d}  \equiv \langle \mathsf{s}^2_{3d} \rangle_{x,y,z}, \\
T  &\equiv&  - \frac{Re}{2} \Big\langle \langle u_i^{3d} u_j^{3d} \rangle_x \dfrac{\p u_i^{2d}}{\p x_j} \Big\rangle_{y,z} \approx -\frac{Re}{2}\Big\langle \langle u^{3d} w^{3d} \rangle_x \dfrac{\p u^{2d}}{\p z} \Big\rangle_{y,z} \slabel{T-definition}
\end{subeqnarray}
Although the transfer term $T$ is defined as the sum of six terms (product of $i=2,3$ by $j=1,2,3$), the approximation in \eqref{T-definition} reflects the experimentally-verified expectation that the dominant contribution comes from the interaction of three-dimensional motions $u^{3d}w^{3d}$ with the vertical shear of the 2D flow $\p_z u^{2d}$ (typically over 90\% of the total in experiments).

\subsection{Sketch and implications for regime transitions} \label{sec:2d-3d-disc}

A sketch of the time-averaged budgets in \eqref{2d-3d-KE} is shown in figure~\ref{fig:energetics-5} (using the fast $t^*$ time scale), which improves on the sketch of figure~\ref{fig:energetics-3}\emph{(b)}. Note that we ignore the vertical buoyancy fluxes $B_z^{2d}$, $B_z^{3d}$ as well as the three-dimensional horizontal buoyancy flux $B_x^{3d}$  since they have been experimentally verified to be negligible (as expected). Panels \emph{(a)} and \emph{(b)} show fluxes of hypothetically different magnitudes under increasing `power throughput' in the system $(Re/2)\langle \Phi^\textrm{adv}_P \rangle_t =  (1/8) \langle Q_m \rangle_t  \theta Re$ (represented by the thickness of the $E\rightarrow P$ arrow). Assuming $\langle Q_m \rangle_t \approx 0.5$, the time- and volume-averaged power throughput in the system is $\theta Re/16$, and  we predict the following:

\begin{figure}
\centering
\includegraphics[width=0.95\textwidth]{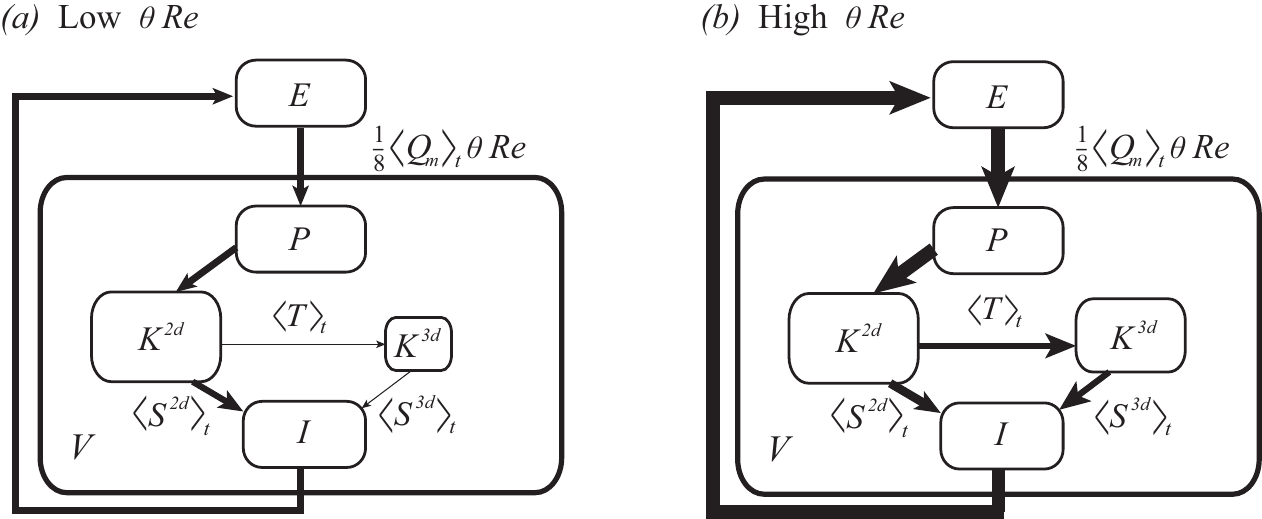}
\caption{Energy budgets of forced flows using the $K=K^{2d}+K^{3d}$ decomposition, refining the budgets of figure~\ref{fig:energetics-3}\emph{(b)}. These budgets in \emph{(a)} and \emph{(b)} only differ in the hypothetical magnitude of the fluxes (with respect to the rescaled time $t^*$), represented by the thickness of the arrows: \emph{(a)} at low $\theta Re$, the power throughput is small and dissipation by $\langle S^{2d}\rangle_t$ is sufficient. \emph{(b)} At high $\theta Re$, the power throughput is high and transfer to $K^{3d}$ by $\langle T\rangle_t$ and dissipation by $\langle S^{3d}\rangle_t$ takes over.}\label{fig:energetics-5}
\end{figure}
\begin{enumerate}
\item for the lowest $\theta \,Re < 20$, the power throughput is $<20/16 = 1.25$, and $\langle S \rangle_t = \langle S^{2d}\rangle_t$ alone is sufficient to dissipate this power via the adjustment of the streamwise velocity profile $u(y,z)$ creating  $O(1)$ gradients $|\p_z u^{2d}|$ and $|\p_y u^{2d}|$. This situation corresponds to the $\LL$ regime, which we have seen in \S~\ref{sec:reg-vis}, is essentially invariant in $x$;

\item for $20 < \theta Re < 50$, the power throughput is $1.25<\langle S\rangle_t <3.12$, and corresponds to the $\HH$ regime, featuring the three-dimensional confined Holmboe waves (CHWs) described in LPZCDL18. To understand the $\LL \rightarrow \HH$ transition, we formulate two distinct hypotheses regarding the energetical importance of CHWs:
\begin{itemize}
\item either HYP-1: the distortion of the two-dimensional flow $u^{2d}$ to yield higher $\p_z u^{2d}, \p_y u^{2d}$ and  $\langle S^{2d} \rangle_t$ `incidentally' renders the flow profile $u^{2d},\, \rho^{2d}$ susceptible to the confined Holmboe instability (CHI) and triggers a transition to a weakly three-dimensional flow state, whose dissipation $\langle S^{3d}\rangle_t$ is insignificant (panel \emph{a}). In other words additional dissipation is achieved primarily by $u^{2d}$ and not by the three-dimensional CHWs, which are simply a by-product of the changes in $u^{2d}$;

\item or HYP-2: the distortion of $u^{2d}$ is no longer sufficient to reach the target dissipation: no two-dimensional solutions exist with the required $\langle S^{2d} \rangle_t$ and the flow \emph{must} `bifurcate' to a three-dimensional state with significant transfer $\langle T\rangle_t$  and additional dissipation $\langle S\rangle_t \gg \langle S^{2d} \rangle_t$ (panel \emph{b}). In other words additional dissipation is achieved by CHWs rather than by a continuing deformation of $u^{2d}$. This hypothesis was expressed in the last sentence of `future direction (ii)' in LPZCDL18 (\S~7.2, p.~540) as a possible mechanism setting the amplitude of Holmboe waves.
\end{itemize}
Experimental data in the next section will allow us to decide which hypothesis is true.

\item for $\theta Re>50$ ($\II$ regime), the power throughput becomes large $>3.12$ and we expect the transfer $\langle T\rangle_t$ and three-dimensional dissipation $\langle S^{3d} \rangle_t$ to be important to close the budgets (panel \emph{(b)}). 
The $\HH\rightarrow \II$ transition may be explained by two hypotheses which are  respectively consistent with those above: 
\begin{itemize}
\item HYP-1: if the CHW is energetically insignificant, its amplitude is presumably not influenced by $\theta Re$. Since it is the two-dimensional flow $u^{2d}$ that responds to $\theta \,Re$, we expect the $\HH\rightarrow \II$ transition to be related to an instability of this base flow;

\item HYP-2: if the CHW is energetically significant in providing three-dimensional dissipation following $\theta \,Re$, its amplitude must be set by $\theta \,Re$ and we thus expect the $\HH\rightarrow \II$ transition to be related to a `secondary' instability of this wave state, perhaps due to a critical (nonlinear) amplitude. 
\end{itemize}

\item for $\theta Re > 100$ (power throughput $ >6.25$) the transition to a sustained $\TT$ regime has a straightforwards explanation: a fully turbulent flow that  sustains high values of $S^{3d}$ in time and space will achieve higher time- and volume-averages of $\langle S^{3d} \rangle_t$ than an intermittently turbulent flow.
\end{enumerate}

\subsection{Experimental validation} \label{sec:2d-3d-experimental-data}

\begin{figure}
\centering
\includegraphics[width=0.98\textwidth]{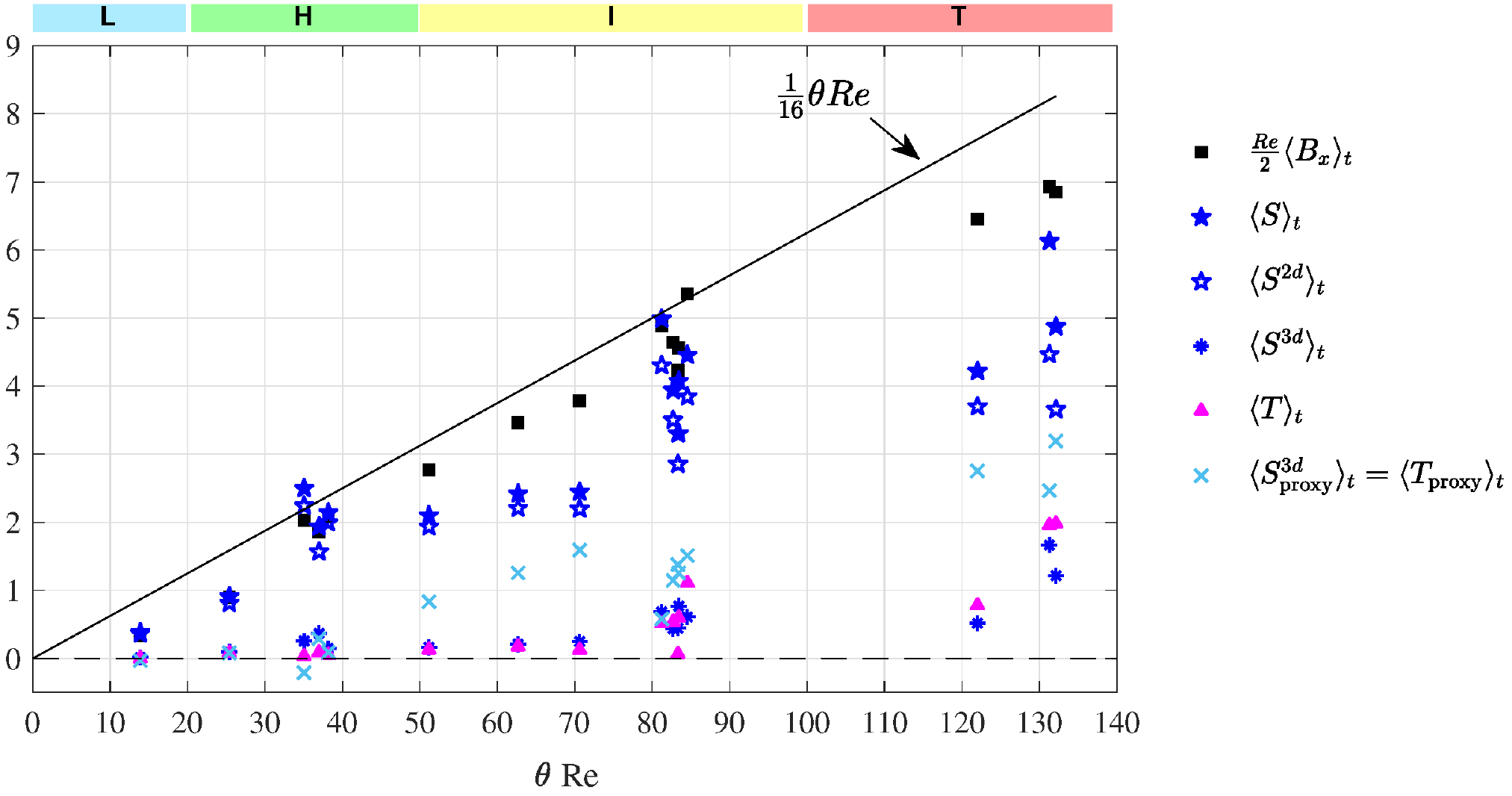}
    \caption{Experimental two-dimensional and three-dimensional kinetic energy budgets in the 16 3D-3C experiments of table~\ref{tab:3d-3c-expts} and figures~\ref{fig:validation},~\ref{fig:dissipation}. The axes, black squares and solid blue stars are identical to those in figure~\ref{fig:dissipation}. The empty blue stars and the blue asterisks show the two-dimensional and three-dimensional decomposition. Magenta triangles represent the rate of transfer of  $K^{2d}$ to $K^{3d}$, and light blue crosses represent the proxy for $\langle S^{3d}\rangle_t$ and $\langle T \rangle_t$ (see \eqref{proxy}). } \label{fig:dissipation-2d-3d-expt}
\end{figure}

We plot the time-averaged fluxes of the $K^{2d},~K^{3d}$  budgets in our 16 3D-3C experiments in figure~\ref{fig:dissipation-2d-3d-expt}. This figure is very similar to figure~\ref{fig:dissipation}, but shows the $S^{2d}+S^{3d}$ decomposition and the transfer term $T$.

We observe that $\langle S^{2d}\rangle_t$ dominates in the $\LL$ and $\HH$ regimes. To mitigate our underestimation of $\langle S^{3d}\rangle_t$ in the $\II$ and $\TT$ regimes  (discussed in \S~\ref{sec:limitations}), we further consider and plot the following trustworthy proxy based on the (verified) steadiness of the kinetic energy reservoirs:
\begin{equation} \label{proxy}
\langle S_{\textrm{proxy}}^{3d} \rangle_{t} = \langle T_{\textrm{proxy}} \rangle_t \equiv \frac{Re}{2}  \langle B_x \rangle_t -  \langle S^{2d} \rangle_{t}, 
\end{equation}
We observe that $\langle S_{\textrm{proxy}}^{3d} \rangle_{t}$ and  $\langle T_{\textrm{proxy}} \rangle_t$ dramatically increase in an approximately linear fashion above the threshold $\theta Re \approx 40$, shortly before the $\HH \rightarrow \II$ transition at $\theta Re = 50$. These observations support the predictions of \S~\ref{sec:2d-3d-disc} and figure~\ref{fig:energetics-5} that the $\II$ and $\TT$ regime correspond to marked increase in three-dimensional dissipation that scales linearly with the power throughput $\theta Re$ due to the upper bound set on the two-dimensional dissipation by hydraulic controls. 

These observations also support HYP-1 in \S~\ref{sec:2d-3d-disc} that Holmboe waves are energetically insignificant and caused by a linear instability triggered by the increased interfacial shear $|\partial_z u|$ reaching a threshold value when $\langle S^{2d}\rangle_t \approx 20$ at the $\LL \rightarrow \HH$ transition (compare the mean profiles between panels\emph{f} and \emph{l} in figure~\ref{fig:regime-snaps-L-H}).  To further support HYP-1, we confirmed that the two-dimensional mean flow in experiment L1 ($\langle u \rangle_{x,t}(y,z)$ and $\langle \rho \rangle_{x,y,t}(z)$) was indeed linearly \emph{stable} to three-dimensional perturbations of the form $\phi' = \hat{\phi}(y,z) \exp (ikx + \sigma t)$ (using the analysis described in LPZCDL18 \S~5.1, which was performed on experiment H4).

\subsection{Spatial structure of energy dissipation}\label{sec:2d-3d-structure}

\begin{figure}
\centering
\includegraphics[width=0.85\textwidth]{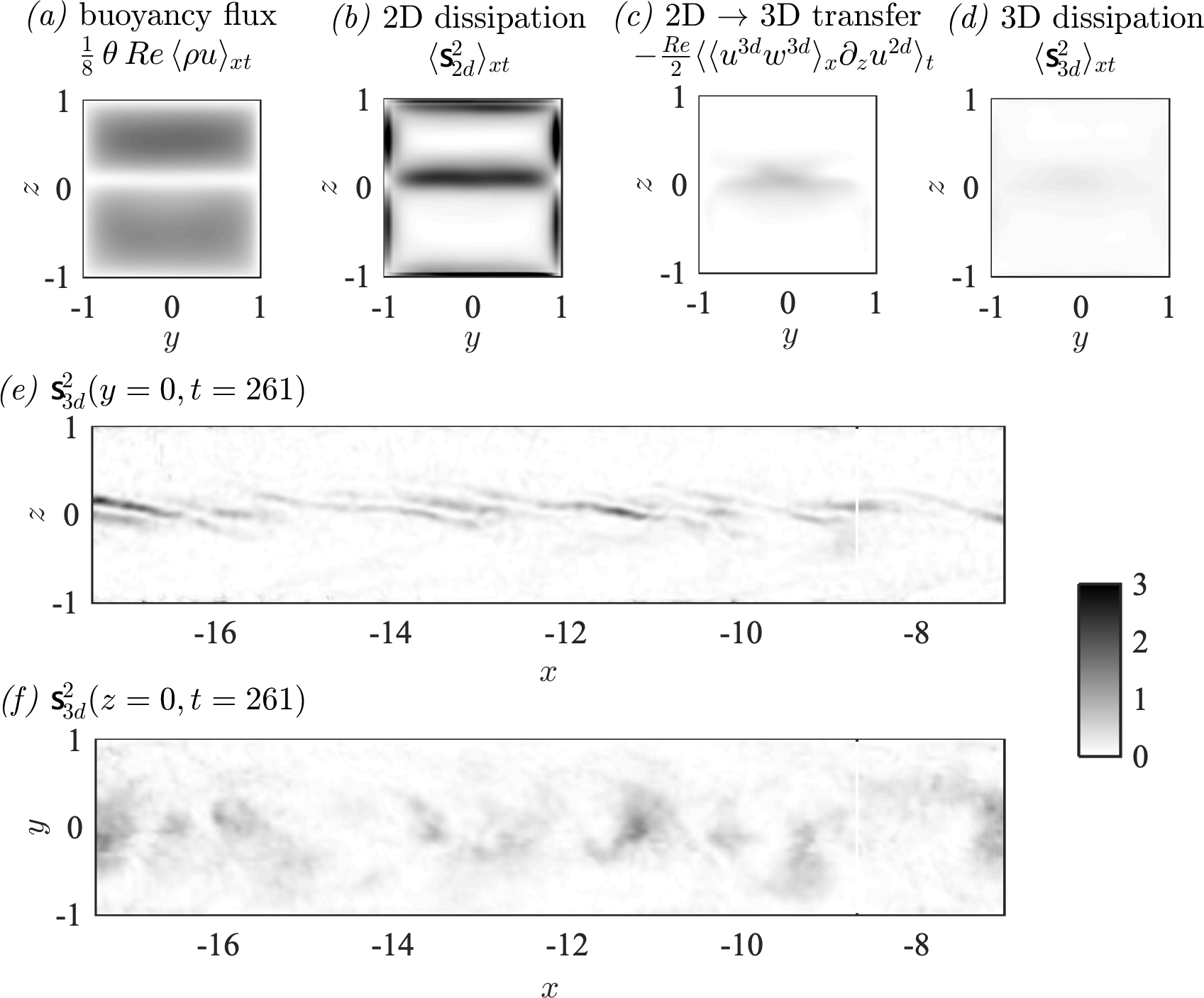}
\caption{Spatial structure of the kinetic energy fluxes in the $\HH$ regime, whose volume-averaged energetics are sketched in figure~\ref{fig:energetics-5}\emph{(a)}.   Two-dimensional cross-sectional structure of the $t$- and $x$-averaged \emph{(a)}  horizontal buoyancy flux; \emph{(b)} two-dimensional dissipation, \emph{(c)} $\mathcal{K}^{2d}\rightarrow \mathcal{K}^{3d}$ transfer; \emph{(d)} three-dimensional dissipation. Instantaneous three-dimensional dissipation in  \emph{(e)} in the vertical mid-plane $y=0$ and \emph{(f)} in the horizontal mid-plane $z=0$. This is the same experiment H1 as in figure~\ref{fig:regime-snaps-L-H}\emph{(a-f)} (instantanenous snapshots are taken at the same time $t=261$). }\label{fig:structure-H}
\end{figure}
\begin{figure}
\centering
\includegraphics[width=0.85\textwidth]{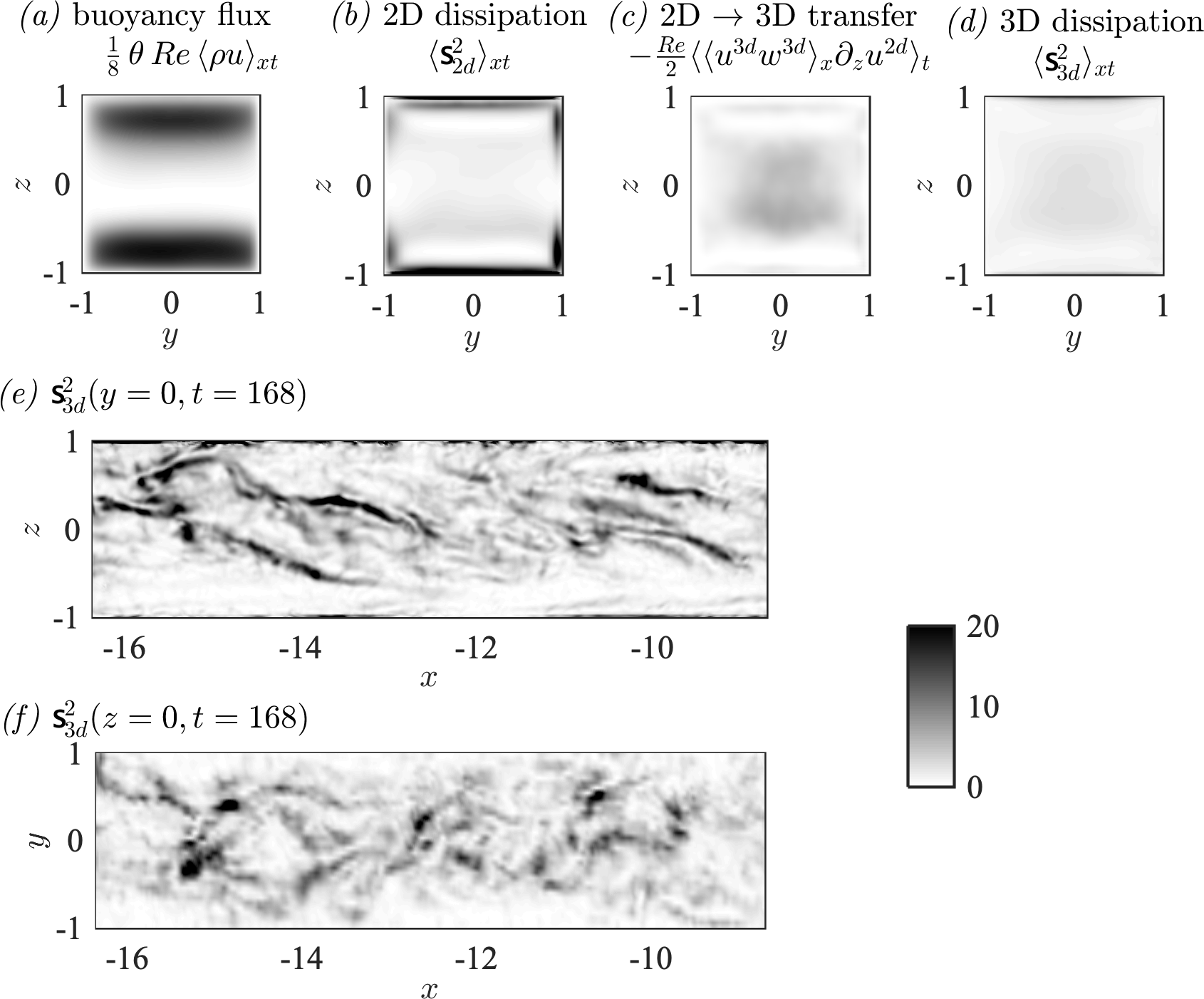}
\caption{Spatial structure of the kinetic energy fluxes in the $\TT$ regime, whose volume-averaged energetics are sketched in figure~\ref{fig:energetics-5}\emph{(b)}. Same panels and legend as figure~\ref{fig:structure-H} for side-by-side comparison. This is the same experiment T2 as in figure~\ref{fig:regime-snaps-I-T}\emph{(g-l)} (instantanenous snapshots are taken at the same time $t=168$).}\label{fig:structure-T}
\end{figure}

In this section, we examine the spatial distribution of energy fluxes to reveal information hitherto hidden by volume averaging. In figures~\ref{fig:structure-H}-\ref{fig:structure-T}, we compare and contrast, for the H1 and T2 experiments respectively, the cross-sectional distribution of the buoyancy flux (panel \emph{a}), two-dimensional dissipation (panel \emph{b}), transfer  (panel \emph{c}) and three-dimensional dissipation (panel \emph{d}).
The cross-sectional average of the data in each panel respectively yields $(Re/2) \langle B_x \rangle_t$, $\langle S^{2d} \rangle_t$, $\langle T \rangle_t$, $\langle S^{3d} \rangle_t$. We also plot instantaneous snapshots of three-dimensional dissipation in the vertical mid-plane plane $y=0$ (panel \emph{e}) and horizontal mid-plane $z=0$ (panel \emph{f}) at the same times as the snapshots in figure~\ref{fig:regime-snaps-L-H}\emph{(a-d)} and figure~\ref{fig:regime-snaps-L-H}\emph{(g-l)}. We recall that the volume-averaged transfer and three-dimensional dissipation are underestimated in the $\TT$ experiment, as can be seen in figure~\ref{fig:dissipation-2d-3d-expt} (next-to-rightmost data series). The proxy data in the latter figure suggests that the (averaged) transfer in figure~\ref{fig:structure-T}\emph{(c)} should be 25\% larger, and the (averaged) dissipation in  figure~\ref{fig:structure-T}\emph{(d-f)} should be 50\% larger.  The time- and volume-averaged power input $(Re/2) \langle B_x \rangle_t$ (which should equal the total $\langle S^{2d} \rangle_t + \langle S^{3d} \rangle_t$) can be read on figure~\ref{fig:dissipation-2d-3d-expt} as $\approx 1$ (H1 experiment) and $\approx 7$ (T2 experiment). Accordingly, the colourbar in figure~\ref{fig:structure-H}-\ref{fig:structure-T} (identical for the all panels of each figure) have respective limits of 3 and 20, equal to about three times the average energy input, allowing for side-by-side comparison of the relative importance of each flux in each regime. Complementary visualisations of slices and averages of the density, velocity and enstrophy fields of experiments H1 (same as in figure~\ref{fig:structure-H}) and T3 (similar to figure~\ref{fig:structure-T}) are available in \cite{partridge_new_2018}.

In both experiments, the power input (panels~\emph{a}) is relatively uniformly distributed within each counter-flowing layer, and low around the sharp interface ($\HH$ regime, figure~\ref{fig:structure-H}) and mixing layer ($\TT$ regime, figure~\ref{fig:structure-T}). In contrast, the two-dimensional dissipation  (panels~\emph{b})) is highly localised at the four duct walls, as well as at the interface in the $\HH$ regime only (in the $\TT$ regime the interfacial shear is comparatively low). The transfer term (panels~\emph{c}) is also highly localised but in the `active core' of the flow, i.e. at the interface ($\HH$) or within the mixing layer ($\TT$). This localised power input of $\mathcal{K}^{3d}$ is then dissipated by three-dimensional motions preferentially in the interior (panels~\emph{d}) as well as a very close to the top and bottom walls in the $\TT$ regime. We also observe that the three-dimensional dissipation is more uniform than the transfer in the cross-section. This suggests complex energy transfer pathways and supports the general conclusion that all the kinetic energy fluxes have very different cross-sectional structures, both in the $\HH$ regime and in the $\TT$ regime. Next, we focus on the instantaneous snapshots of three-dimensional dissipation in panels~\emph{(e-f)}. Beyond the observation that its volume-average $S^{3d}$ is only significant in the $\TT$ regime, we see, without surprise, that its spatial structure is highly heterogeneous.  `Wispy' regions with considerable three-dimensional structure feature much enhanced dissipation, several times larger than their respective volume-average, especially in the $\TT$ regime where it locally exceeds the limit of colour bar.

\section{Conclusions} \label{sec:ccl}

\subsection{Summary}

In this paper, we investigated the transition in the long-term qualitative behaviour, or flow regime, of geophysically-relevant sustained stratified shear flows as two key forcing parameters are varied. We performed laboratory experiments in the Stratified Inclined Duct (SID) setup (figure~\ref{fig:setup}) which features four qualitatively different regimes: laminar ($\LL$), Holmboe waves ($\HH$), intermittently turbulent ($\II$) and fully turbulent ($\TT$), with increasing three-dimensionality and mixing intensity (figures~\ref{fig:regime-snaps-L-H}-\ref{fig:regime-snaps-I-T} and table~\ref{tab:regimes}). These regimes occupy distinct regions in the two-dimensional space of non-dimensional input parameters: duct tilt angle $\theta \in [-1^\circ,6^\circ]$ and Reynolds number $Re\in [300,5000]$ (figure~\ref{fig:regime-data-input-params}). Although these regimes have been observed since at least \cite{macagno_interfacial_1961}, we argued that previous attempts to explain the transitions were unsuccessful.  \cite{meyer_stratified_2014} (ML14) recognised the importance of both $\theta$ or $Re$ and proposed a heuristic scaling of iso-regime curves scaling with the nondimensional group  $\theta Re^2$. However, this scaling does not agree with our regime diagram obtained in a smaller duct (figure~\ref{fig:regime-data-input-params}) and motivated our search for a scaling law resting on a firm physical basis and backed by experimental data.

Therefore, we derived, from first principles, evolution equations for the volume-averaged potential and kinetic energy in a control volume of arbitrary length, whose cross-section is bounded by the four walls of our square duct (equations \eqref{dKdt-final},~\eqref{dPdt-final}, sketched in figure~\ref{fig:energetics-1}). We then introduced a simplified two-layer frictional hydraulics model (figure~\ref{fig:energetics-2}) to make modelling progress and simplify the energy budgets in SID flows. We distinguished between, on one hand, `lazy flows' at low $|\theta| \lesssim 2^\circ$, in which the forcing is primarily hydrostatic and dwarfed by viscous friction; and on the other hand, `forced flows', at high $\theta \gtrsim 2^\circ$, in which the forcing is primarily gravitational and is closely balanced by viscous friction (figure~\ref{fig:lazy-forced}). We showed that these flows have different energetics (figure~\ref{fig:energetics-3}) and that, in a statistically-steady sense (averaged over sufficiently long times), any control volume of a forced flow exhibits remarkably simple energy budgets characterised by a single potential power input from the exterior, a single potential-to-kinetic conversion power and a single kinetic dissipation power, all equal in magnitude (equation \eqref{steady-state-dissipation} and figure~\ref{fig:energetics-3}\emph{(b)}). This led us to propose the physical hypothesis that regime transitions are caused by increasing values of the suitably-rescaled time- and volume-averaged rate of kinetic energy dissipation, or square norm of the strain rate tensor $\langle \mathsf{s}_{ij}\mathsf{s}_{ij} \rangle_{x,y,z,t}$ (equations~\eqref{def-S} and \eqref{reg-s2}), and to deduce the `forced flow' corollary that regime transitions should therefore scale like $\theta Re$.

We validated this theory in two ways. First, our experimental regime diagram (figure~\ref{fig:regime-mSID-resintheta}) confirmed the $\theta Re$ scaling predicted by the corollary. Second, we obtained a comprehensive data set of unprecedented volumetric measurements of the density and three-component velocity fields in 16 experiments spanning all four regimes (table~\ref{tab:3d-3c-expts} and figures~\ref{fig:validation},~\ref{fig:dissipation}). Our time- and volume-averaged measurements of all energy fluxes confidently support our theoretical `forced flow' energy budget model, as well as the above physical hypothesis, despite the experimental challenges of obtaining accurate kinetic energy dissipation rates (figure~\ref{fig:resolution}).

We delved deeper into the above hypothesis by deriving  budgets for the two-dimensional (streamwise-invariant) and three-dimensional components of kinetic energy for forced flows (equation \eqref{dK2d3ddt-final}). We further hypothesised that flows with low power-throughput and thus low dissipation power (low $\theta Re$, figure~\ref{fig:energetics-5}\emph{(a)}) may be able to dissipate energy exclusively two-dimensionally by increasing the magnitude of their exchange flow rate (volume flux) and their streamwise-invariant wall and interfacial shear ($\LL$ and $\HH$ regimes). By contrast, flows with high power-throughput (high $\theta Re$, figure~\ref{fig:energetics-5}\emph{(b)}) are not be able to dissipate enough energy two-dimensionally due to the upper limit on the exchange flow rate set by hydraulic controls, and thus have to transition to intermittently and fully turbulent regimes with increasingly three-dimensional dissipation scaling with $\theta Re$. We validated this hypothesis with our volumetric experimental data set (figure~\ref{fig:dissipation-2d-3d-expt}) despite having to use indirect evidence (equation \eqref{proxy}) to mitigate the experimental under-estimation of three-dimensional dissipation. Based on further observations, we suggested that \emph{(i)} the $\LL \rightarrow \HH$ transition  was caused by a Holmboe instability triggered by the increasing interfacial shear resulting from the two-dimensional dissipation scaling with $\theta Re$; \emph{(ii)} the $\HH \rightarrow \II$ transition might be triggered by another primary instability of the base flow rather than by ever-growing Holmboe waves since the latter are energetically insignificant. We also showed that energy transfers in the three-dimensional experimental volume were complex and heterogeneous in space, particularly in the more turbulent regimes (figures~\ref{fig:structure-H} and \ref{fig:structure-T}).

To conclude, we believe that we have achieved our initial aim, since our results provide the first mutually-consistent physical basis and experimental data to explain the observed transitions in the qualitatively different long-term dynamics of SID flows. The generality of these results provides a useful basis for the study of a broader range of sustained stratified shear flows found in Nature.

\subsection{Unanswered questions}

Our results raise at least four unanswered questions:

\begin{enumerate}

\item What is the dynamical explanation for the $\II \rightarrow \TT$ transition? We proposed that the $\LL \rightarrow \HH$ and $\HH \rightarrow \II$ transitions were caused by stratified shear instabilities resulting from modifications in the parallel base flow slaved to the energy throughput $\theta Re$. We explained that, energetically, sustained turbulence in the $\TT$ regime was able to achieve higher time-averaged three-dimensional dissipation than intermittently in the $\II$ regime. However, does this transition occur by a gradual lengthening of the period of turbulent events with respect to laminar events or by a more abrupt bifurcation? In other words, do `intermediate' solutions exist with a range of turbulent/laminar period ratios or a range of different dissipative structures?  The dynamical details of the transition between intermittency and sustained turbulence, and the quantitative explanation for the transition occurring at $\theta Re \approx 50$ remain open questions.

\item How to explain flow regime transitions in horizontal ducts or duct inclined at a slightly negative angle? We indeed observed Holmboe waves and intermittent turbulence for $\theta = 0^\circ$ (figure~\ref{fig:regime-data-input-params}), yet our forced flow scaling of transitions with $\theta Re$  only applies for $\theta \gtrsim \alpha$ (we recall that $\alpha \equiv H/L$ is the inverse aspect ratio of the duct, see \eqref{definition-alpha}). Flows at $|\theta| \lesssim \alpha$ have more complex energetics (figure~\ref{fig:energetics-3}\emph{(a)}), and we have seen that, in such flows, transitions appear to scale with $\theta Re^2$ instead of $\theta Re$ (figure~\ref{fig:regime-mSID-resintheta}).  Further work is needed to understand lazy flow dynamics and explain this $\theta Re^2$ scaling.

\item Why did ML14 observe a different transition scaling in a different duct geometry? As evidenced by the dashed line in figure~\ref{fig:regime-data-input-params} and as discussed in \S~\ref{sec:ML14}, their experiments in a larger (but still square) duct ($H=100$~mm \emph{vs} $45$~mm in this paper) suggested a $\theta Re^2$ scaling (both for lazy and forced flows) in disagreement with our theory. However, we note that the Reynolds numbers in ML14 are typically larger than ours. At sufficiently large $Re$, wall boundary layers are not fully-developed and do not span the whole cross-section of the duct as was typically the case in the data shown in this paper. Instead, wall boundary layers become sufficiently thin that the volume-averaged contribution of wall dissipation is no longer of order $1$ but scales with $Re^{1/2}$. This apparently undermines our simple hypothesis \eqref{reg-s2} that increasingly turbulent regimes correspond to increasing values of the \emph{volume-averaged} dissipation well above `laminar' $O(1)$ values but more work is required to investigate this question. 

\item What is the role of mixing? In this paper, we focused on kinetic energy dissipation to explain regime transitions and did not explicitly derive or represent irreversible mixing in the energy budgets. Irreversible mixing is implicitly accounted for in the mass flux $Q_m$ \eqref{Qm-def}, to which the energy throughput of forced flows is proportional (see \eqref{s2-Resintheta}). Although the black contours in figure~\ref{fig:regime-mSID-resintheta} show that the mass flux has a complicated $Q_m(\theta, Re)$ dependence (due primarily to the volume flux $Q(\theta, Re)$ and secondarily to mixing), we made the reasonable assumption that, in forced flows, $Q \approx Q_m \approx 0.5$ (leading to \eqref{s2-Resintheta-simplified}). We believe that neglecting mixing in this fashion is acceptable for the work in this paper, but acknowledge that a better understanding of the $Q(\theta, Re)$ and $Q_m(\theta, Re)$ relations is desirable. More generally, beyond the $Q_m/Q$ question and its (moderate) impact for the energy throughput in forced flow, we believe that the study of mixing and mixing efficiency in sustained stratified shear flows remains a major research objective. However we are currently not able to measure mixing accurately in experiments; the Batchelor lenghtscale is typically $Sc^{1/2} \approx 25$ times smaller that the Kolmogorov scale, which is already challenging to resolve (\S~\ref{sec:limitations}).
For a more detailed discussion about mixing in the SID experiment, including an explicit representation of irreversible mixing in energy budgets, see \cite{lefauve_waves_2018} \S~6.7. 


\end{enumerate}

\vspace{0.5cm}

\noindent \textbf{Acknowledgements}

\vspace{0.2cm}

AL is funded by an Engineering and Physical Sciences Research Council (EPSRC) Doctoral Prize. All authors acknowledge funding from the EPSRC under the Programme Grant EP/K034529/1 `Mathematical Underpinnings of Stratified Turbulence' (MUST), and from the European Research Council (ERC) under the European Union's Horizon 2020 research and innovation programme under grant No 742480  `Stratified Turbulence And Mixing Processes' (STAMP). We thank the `MUST team' in DAMTP for helpful discussions and especially Prof. Colm Caulfield about the two-dimensional/three-dimensional decomposition. We are grateful for the invaluable experimental support of Prof. Stuart Dalziel and of the technicians of the G. K. Batchelor Laboratory.

\appendix{
\section{Experimental constraints} \label{sec:appendix-exp}

The physical constraints currently limiting the resolution and temporal duration of our experimental measurements are as follows:
\begin{itemize}
\item the streamwise and vertical resolutions $\Delta x \equiv \ell/n_x, \ \Delta z \equiv 2/n_z$ (where $n_x, n_z$ are the number of sPIV vectors in each direction) are generally equal and limited by the resolution of the cameras, the size of the PIV particles (typically $50~\mu$m) and their seeding density. Using 8~MPixel cameras, $31\times31$~Pixel interrogation windows, a 75~\% overlap, and volumes of length $\ell \approx 11$, we typically obtained $n_x\approx 500,\, n_z  \approx 100$, i.e. $\Delta x \approx \Delta z \approx 0.02$. 
Density data were obtained at higher resolution because of the absence of interrogation windows in PLIF, but since this higher resolution was not needed for the analysis in this paper, they were smoothed before being interpolated onto the grid of the velocity data;

\item the spanwise resolution $\Delta y \equiv 2/n_y$  is limited by the finite thickness of our laser sheet (required for sPIV measurements) estimated to be $\approx 1.5$~mm $\approx H/30$, dictating $n_y\approx 30-40$ as a good compromise to avoid excessive redundancy of overlapping laser sheets, and therefore a typical resolution $\Delta y \approx 0.05-0.07$ (coarser than $\Delta x,\, \Delta z$);

\item the temporal resolution $\Delta t\equiv n_y \delta t$ of our measurements is primarily limited by the previously set $n_y$ and the laser frequency of $\delta t^{-1}$ (a maximum of $100$~Hz in dimensional units, i.e. 100 \emph{double} pulses per second). This results in a typical non-dimensional lower bound $\Delta t \gtrsim 30 \times 100^{-1} \times \Delta U / H = (1.2 \nu/H^2)Re \approx  Re/1600$, making the near-instantaneous `freezing' of volumes better (i.e. $\Delta t$ smaller) in low-$Re$ flows than in high-$Re$ flows (for a given $H$ and $\nu$). For the flows considered in this paper, $\Delta t\approx  1-4$~ATU  (the lower bound $\Delta t \approx Re/1600$ was only rarely realised since the laser could only be set at its maximum frequency for the fastest, highest-$Re$ flows). 

\item the duration of the recorded data, $\tau \equiv n_t \Delta t$, and therefore the number of successive volumes measured $n_t$, is limited by the available RAM storage memory (50~GB) dedicated to each camera (two cameras for sPIV and one camera for PLIF). A total of 150~GB of raw data typically yielded $\approx 18000$ frames per camera, i.e. $\approx 9000$ sPIV fields or $n_t=9000/n_y\approx 300$ volumes spanning a duration $\tau \approx 10^2-10^3$~ATU (typically a few minutes). Although $\tau$ is typically shorter than the maximum duration of an experiment (before the flooding of the controls, determined by the size of the reservoirs), we refer to it as the `duration of an experiment' in this paper for simplicity.
\end{itemize}

\section{Estimation of energy fluxes} \label{sec:estimation-fluxes}

Based on the two-layer hydraulic model of figure~\ref{fig:energetics-2}, we use the definitions for the energy fluxes in the $K$ and $P$ budgets \eqref{K-boundary-fluxes}, \eqref{K-volume-fluxes}, \eqref{P-boundary-fluxes} to estimate the following (derivations can be found in L18,~\S~6.3.1):

\begin{itemize}
\item the advective boundary flux $K$ is
\begin{equation}\label{PhiKadv}
\Phi_K^\textrm{adv} = -\frac{Q^3}{\ell}\Big\{ \frac{\eta_L}{(1-\eta_L^2)^2} -  \frac{\eta_R}{(1-\eta_R^2)^2} \Big\} \leq 0 \quad \textrm{since} \quad \eta_L \geq \eta_R,
\end{equation}
it is thus always negative (it acts as a sink to $K$) since the interface must slope down. In other words, the inflow of kinetic energy in $V$ by the velocities $u_{1L}$, $u_{2R}$ is always smaller than the outflow by the velocities $u_{2L}$, $u_{1R}$. (Note that even more negative $\Phi_K^\textrm{adv}$ would be obtained by relaxing the assumption of uniform flow in each layer and taking into account the non-unitary velocity distribution coefficient when evaluating $\langle u^3 \rangle_{y,z}$, which is typically greater for the thin outflowing layers than for the thick inflowing layers).

Importantly, we note that $\Phi_K^{\textrm{adv}} =0$ if $V$ is approximately periodic in $x$, i.e. if velocities and interface position are identical at the left and right boundaries. For any general $V$, this requires that the interface is flat everywhere $\eta(x) =0$, which as explained in \S~\ref{sec:two-layer-model} corresponds to forced flows guaranteed at large tilt angles $\theta>\alpha$.

\item the pressure boundary flux of $K$ is
\begin{equation} \label{PhiKpre}
\Phi_K^\textrm{pre} =  \frac{1}{4\ell} \langle u (\eta - z) \rangle_{y,z}|_{L-R}   =  0,
\end{equation}
under the assumptions of no barotropic flow $\langle u \rangle_{x,y,z}=0$  and of hydrostatic flow (in particular that $u$ does not depend on $z$). We will therefore neglect this flux.

\item  the viscous boundary flux of $K$ is
\begin{equation}
\Phi_K^\textrm{vis}  =\frac{8Q^2}{\ell \, Re} \Big\{ \frac{\eta_L \eta'_L}{(1-\eta_L^2)^2} - \frac{\eta_R \eta'_R}{(1-\eta_R^2)^2} \Big\},
\end{equation}
We note that, similarly to the advective flux, $\Phi_K^\textrm{vis}=0$  in forced flows (i.e. if $V$ is periodic). However, for the large $Re\gg 1$ investigated here, this flux will be neglected compared to the advective flux $|\Phi_K^\textrm{vis}|\ll |\Phi_K^\textrm{adv}|$.

\item  the advective boundary flux of $P$ is
\begin{equation} \label{PhiPadv}
\Phi_P^\textrm{adv} =  \underbrace{\frac{1}{4\ell}Q_m\Big( \frac{\eta_L}{1-\eta_L^2}-\frac{\eta_R}{1-\eta_R^2} \Big)}_{\substack{\textrm{hydrostatic} \\ \textrm{forcing}}\ >0} \quad + \underbrace{\vphantom{\frac{\eta_L}{1-\eta_L^2}} \frac{1}{4} Q_m \, \theta}_{\substack{\textrm{gravitational} \\ \textrm{forcing}}\ >0}  > 0.
\end{equation}
We note that $\Phi_P^\textrm{adv}$  has two distinct positive components: hydrostatic forcing and gravitational forcing, as already identified in \eqref{x-momentum}. Consistently with the discussion of \S~\ref{sec:two-layer-model}, we see here that for forced flows the hydrostatic term cancels and \emph{only the gravitational forcing remains}.

\item the diffusive boundary flux of $P$: 
\begin{equation}
\Phi_P^\textrm{dif} = \frac{1}{4\ell \, Re \, Sc} \Big[   \theta \big\{  \big(\bar{x}+\frac{\ell}{2}\big)\eta'_R - \big(\bar{x}-\frac{\ell}{2} \big)\eta'_L \big\} +  (\eta_L \eta'_L - \eta_R\eta'_R ) \Big],  
\end{equation}
where again, $\Phi_P^\textrm{dif}=0$ for forced flows. Moreover, just like $\Phi^\textrm{vis}_K$, we neglect this flux for the large $Re$ and $Sc$ used here since $|\Phi_P^\textrm{dif}|\ll|\Phi_P^\textrm{adv}|$.

\item the horizontal buoyancy flux:
\begin{equation} \label{Bx}
B_x = \frac{1}{4}  \langle \rho u \rangle_{x,y,z} \, \theta = \frac{1}{4}Q_m  \,\theta >0,
\end{equation}
which is \emph{exactly equal to the gravitational component of} $\Phi^\textrm{adv} _P$ (see \eqref{PhiPadv}).

\item the vertical boundary flux: 
\begin{equation} \label{Bz}
B_z =  -Q_m \frac{\eta_L - \eta_R}{4\ell} \le 0,
\end{equation}
under the assumption that the center of mass of a slab of dense ($\rho=1$) fluid drops by $\eta_L-\eta_R$ over the length $\ell$ (i.e. it has a negative vertical velocity), and conversely for a slab of buoyant ($\rho=-1$) fluid.
In the absence of any other vertical motion other than those consistent with hydraulic theory, it is thus negative, meaning that it acts as a source term for $K$ (where it appears as $-B_z$, see \eqref{dKdt-final}) and as a sink for $P$ (where it appears as $+B_z$, see \eqref{dPdt-final}). We note that this flux also cancels for forced flows. However, if we relax the hydraulic assumptions (as will be required to investigate the laboratory flows in this paper), non-trivial vertical motions (turbulence) may render $B_z$ sign-indefinite. We therefore consider this flux to be \emph{irreversible} (negative-definite) in flows close to the hydraulic assumptions ($\LL$ and $\HH$ regimes) and potentially \emph{reversible} (sign-indefinite) in flows where vertical motions may be large ($\II$ and $\TT$ regimes).

\item the conversion of $I$ to $P$:
\begin{equation}
\Phi_P^\textrm{int} = \frac{1}{4 \, Re \, Sc} \Big\{  - \underbrace{  \frac{\eta_L - \eta_R}{\ell} \, \theta }_{> \, 0 \  \textrm{and} \  \ll \, 1}  + 1 \Big\} \approx \frac{1}{4 \, Re \ Sc}  ,
\end{equation}
since $\langle \rho \rangle_{y,z} = \eta$, by definition of $\eta$, assuming collocation of the velocity and density interfaces, and $\langle \rho \rangle_{x,y}|_{B-T}=1-(-1)=2$. Given the large $Re$ and $Sc$ investigated here, we neglect it.

\item the viscous dissipation: under the assumptions of hydraulic theory, $D=0$. When relaxing these assumptions, as we will do shortly, $D>0$ but is unknown.  We show in \S~\ref{sec:implications} that it can be deduced in the simplified budget of forced flows.

\end{itemize}

}

\bibliographystyle{jfm}
\bibliography{references}

\end{document}